\documentclass[11pt,onecolumn]{IEEEtran}
\usepackage{amsfonts,amssymb}
\usepackage{latexsym,amscd,amsmath}
\usepackage{amsbsy}
\usepackage{indentfirst,mathrsfs}
\usepackage{array}
\usepackage{color}
\usepackage{multirow}
\usepackage{makecell}

\normalsize
\newcommand{\Tr}{{\rm Tr}}
\numberwithin{equation}{section}
\newtheorem{thm}{Theorem}
\newtheorem{lem}{Lemma}

\newtheorem{prop}{Proposition}

\newtheorem{remark}{Remark}
\newtheorem{problem}{Problem}
\begin{document}
\title{Infinite families of optimal and minimal codes over rings using simplicial complexes}
\author{Yanan Wu,
\thanks{Y. Wu and Y. Pan are with the Key Laboratory of Mathematics Mechanization, Academy of Mathematics and Systems Science,
Chinese Academy of Sciences, Beijing 100190, China. Email: yananwu@amss.ac.cn, panyanbin@amss.ac.cn}
 Tingting Pang, \thanks{T. Pang is with the School of Information Science and Engineering, Shandong Normal University, Jinan 250358, China. Email: pangtingting@sdnu.edu.cn}
Nian Li, \thanks{N. Li is with the Hubei Provincial Engineering Research Center of Intelligent Connected Vehicle Network Security,  School of Cyber Science and Technology, Hubei University, Wuhan 430062, China. Email: nian.li@hubu.edu.cn}
Yanbin Pan, Xiangyong Zeng\thanks{ X. Zeng is with Hubei Key Laboratory of Applied Mathematics, Faculty of Mathematics and Statistics, Hubei University, Wuhan 430062, China. Email:  xzeng@hubu.edu.cn}}
\date{}
\maketitle
\begin{quote}
{\small {\bf Abstract:}} In this paper, several infinite families of codes over the extension  of non-unital non-commutative rings are constructed  utilizing general simplicial complexes. Thanks to the special structure of the defining
sets, the principal parameters of these codes are characterized.  Specially, when the employed simplicial complexes are generated by  a single maximal element, we determine their  Lee weight distributions completely.  Furthermore, by considering the Gray image codes and the corresponding subfield-like codes, numerous of linear codes over $\mathbb{F}_q$ are also obtained, where $q$ is a prime power. Certain conditions are given to ensure the above linear codes  are (Hermitian) self-orthogonal  in the case of  $q=2,3,4$. It is noteworthy that  most of the derived  codes  over $\mathbb{F}_q$ satisfy the Ashikhmin-Barg's condition for minimality. Besides, we obtain two infinite families of distance-optimal codes over $\mathbb{F}_q$ with respect to the Griesmer bound.


\end{quote}

\section{Introduction}

\subsection{Background}
The study of codes over rings has garnered significant interest within the domain of algebraic coding theory in recent times.  The origins of this research can be traced back to the 1970s.   In 1994, Hammons et al. \cite{HKC} constructed some quaternary linear codes and found some well-known nonlinear binary codes  by studying linear codes over the ring $\mathbb{Z}_4$ under the Gray map.  Subsequently, many scholars  considered  various classes of constacyclic codes over finite commutative rings with their Gray images,  see  (\cite{CC,D,DKM,KL,SQS}) for more information.  Determining the Lee weight distribution of a code over rings  is an important research topic in coding theory, because it does give vital information of both practical and theoretical significance. However, explicitly characterizing  the Lee weight distribution is usually a challenging task. Consequently, researchers have mainly concentrated on cases where particular rings or defining sets are meticulously chosen.  For instance,   selecting all nonzero elements of  $\mathbb{F}_{2^m} +u\mathbb{F}_{2^m}$ as a defining set led to a class of two-weight trace codes over $\mathbb{F}_{2^m} +u\mathbb{F}_{2^m}$ with $u^2=0$  \cite{SLS}.
Using the defining set $\mathcal{Q}+u\mathbb{F}_{p^m}$,   Shi and Wu et al. \cite{SWLS},  and Shi and Guan et al.  \cite{SGS} presented several classes of few-weight trace codes over  $\mathbb{F}_{p} +u\mathbb{F}_{p}$ in the case of  $u^2=0$ and $u^2=u$, respectively, where $p$ is an odd prime and  $\mathcal{Q}$ is the set of all square elements of $\mathbb{F}_{p^m}^*$. Additionally, Liu and Maouche \cite{LM}  constructed a class of two or few-weight trace codes over $\mathbb{F}_q+u\mathbb{F}_q$ by virtue of the cyclotomic class in $\mathbb{F}_q$, where $u^2=0$ and $q$ is a prime power.
The investigation on minimal codes  is another hot  topic  nowadays because of  their rich applications in secret sharing
schemes \cite{JC}, graph theory,  secure two-party computation, and they could be decoded with a minimum distance decoding method  \cite{AB}.

Recently, Chang and Hyun \cite{CH} introduced a novel method of constructing binary linear codes by using  simplicial complexes. Applying this method, they constructed  a  family of binary minimal linear codes over finite fields. In 2020, Hyun et al. \cite{HLL}   presented infinite families of optimal binary linear
codes derived from general simplicial complexes of $\mathbb{F}_2^m$  via the defining-set approach. Motivated by their works, many researchers attempted to construct optimal or good linear codes by using  simplicial complexes, which have  only one  or two maximal elements. Some relevant  results are documented in \cite{LS,PL,SL,WLX}.
In 2023, Hu and Xu et al. \cite{HY} obtained more infinite families of optimal codes over $\mathbb{F}_q$ from  general simplicial complexes of $\mathbb{F}_q^m$ for any prime power $q$, which totally extended the results of  \cite{HLL} from $\mathbb{F}_2$ to $\mathbb{F}_q$.

The utilization of simplicial complexes as a framework for constructing linear codes over finite fields has been recognized as a highly effective strategy. This insight has inspired a wave of academic enthusiasm, prompting researchers to extend this innovative technique to the domain of ring-based code construction. In  \cite{WZY}, Wu, Zhu and Yue  defined two classes of codes over the ring $\mathbb{F}_2+u\mathbb{F}_2$  with $u^2=0$ by using  the standard inner product and two suitable defining sets consisting of  simplicial complexes with one maximal element. Via the Gray map, optimal codes with respect to the Griesmer bound could be obtained.  Shortly thereafter, Shi and Li  \cite{SL2} studied the codes over a nonchain ring also by virtue of  simplicial complexes with one maximal element. They found that some minimal and new distance-optimal  binary codes with few-weight could be produced. When simplicial complexes were generated by two maximal elements, Wu and Li et al. \cite{WLZX} constructed two infinite families of four-Lee-weight quaternary codes. Furthermore, under the Gray map, they produced two infinite families of binary nonlinear codes and one infinite family of binary minimal linear codes.   Notably, the pioneering attempt to study the structure of codes over the non-unital non-commutative ring using simplicial complexes is attributed to  Sagar and Sarma \cite{SS}. They primarily concentrated on the codes over the ring of size four with the help of certain simplicial complexes generated by a single maximal element.  Self-orthogonal and minimal codes were derived from their constructions. Moreover, they obtained a couple of optimal codes meeting the Griesmer bound.
\subsection{Motivation}
As evident from previous works, simplicial complexes have been proven  to be a powerful tool for constructing families of codes over rings which may have good performance under the Gray map. Generally, it is a  difficult work to calculate the Lee weight distribution, and thus,  almost all the known studies have focused on rings of order $p^2$,  by making use of  simplicial complexes generated by no more than two maximal elements, where $p$ is a prime. Therefore, we pose the following questions:
\begin{itemize}
  \item Is it feasible to devise a universal methodology for computing the principal parameters of codes over rings of order $q^2$, particularly when contemplating the general simplicial complexes within the vector space $\mathbb{F}^m_q$?
   Here $q$ is any prime power and ``general simplicial complexes" means that the simplicial complexes generated by arbitrary number of maximal elements.

  \item Is it within our capability to delineate the intrinsic structures of the codes we derive, including attributes such as optimality, self-orthogonality, and minimality?  Further, can  new  codes   be produced when employing general simplicial complexes?
\end{itemize}
%
%

\subsection{Our Contributions}
In \cite{Fine}, the authors classified all the rings of size $p^2$ into 11 inequivalent rings, where $p$ is a prime. Among the 11 rings, there are only two of them are  non-unital non-commutative rings (namely, $E$ and $F$). Let $s$ be any positive integer, $q=p^s$ and $\mathcal{R}$ be an extension of $E$ (or $F$)  of degree $s$.  Based on the questions above, we mainly investigate the codes over the ring $\mathcal{R}$ by selecting  suitable defining sets,  which are composed of simplicial complexes generated by any number of maximal elements. The main contributions in this paper are the following.

\begin{itemize}
  \item Utilizing the orthogonality inherent in inner products and the fundamental principle of inclusion-exclusion, we provide a general method to determine the principal parameters  of these codes over  $\mathcal{R}$.
      This method can be applied to generalize  the previous results on the  codes over rings using simplicial complexes of $\mathbb{F}_q^m$ from $q=2$ to any prime power $q$.

  \item Discover the Lee weight distributions of codes over the rings $E^s$ and $F^s$ as we explore simplicial complexes generated by a single maximal element in \eqref{Ring-Trace-Code}, detailed in Propositions  \ref{thm3} and \ref{thm4}. Accordingly, several classes of few-Lee-weight codes including two-Lee-weight, three-Lee-weight and five-Lee-weight codes can be produced.  Notably, these two propositions not only  illustrate our main results, but also extend the work of  \cite{SS}.


  \item  Leveraging the Gray image  of the code over $\mathcal{R}=F^s$ and the subfield-like codes, we uncover  numerous infinite families of minimal codes over $\mathbb{F}_q$. Additionally,  for $q=2$,  $3$ and  4,  the sufficient conditions are provided to guarantee the self-orthogonality of the Gray image codes and the subfield-like codes derived from our construction, respectively.
It is noteworthy that several infinite families of distance-optimal linear codes with respect to the Griesmer bound over $\mathbb{F}_q$ are also obtained.


%

   \item To our best knowledge, we are first to contemplate the application of general simplicial complexes to the domain of codes over extended rings of arbitrary degree. Compared with known results, the codes obtained in this work have more flexible and new parameters, see Table \ref{Known codes}.
\end{itemize}

\section{Definitions and Preliminaries}
Let $q$ be a prime power. In this section, we recall some notation about Lee weight, simplicial complexes and the known generic construction of linear codes. Further, we propose our construction of codes over rings and give the corresponding Lee weight expression of codewords.
\subsection{Lee Weight and Gray Map}
Let $p$ be a prime  and $E,\,F$
be the rings of order $p^2$ given in \cite{Fine}, where
$$E=\langle a,b\,|\,pa=pb=0,a^2=a,b^2=b,ab=a,ba=b\rangle,$$
$$F=\langle a,b\,|\,pa=pb=0,a^2=a,b^2=b,ab=b,ba=a\rangle.$$
It has been pointed out in  \cite{Fine} that $E,\,F$ are both non-unital  non-commutative rings. Moreover, $F$ is the opposite ring of $E$. Suppose $q=p^s$ and $\mathcal{R}$ is an extension of $E$ (or $F$)  of degree $s$, i.e.,  $\mathcal{R}=E^s$ (or $F^s$), each element of $\mathcal{R}$ can be expressed as $ax+cy$ for some $x,y\in \mathbb{F}_q$, where $c=a-b$. Further, according to the presentations, let $r_1,r_2\in \mathcal{R}$, then there exist $(x_1,y_1),(x_2,y_2)\in\mathbb{F}_{q}^2$ satisfying $r_1=ax_1+cy_1$, $r_2=ax_2+cy_2$. Thus, the multiplication operation between $r_1$ and $r_2$ is defined by
\begin{eqnarray}\label{ring-multiplication}
r_1 r_2=(ax_1+cy_1)(ax_2+cy_2)=\left\{\begin{array}{ll}
 ax_1 x_2+cx_2y_1, & {\rm if}\; \mathcal{R}=E^s;\\[0.05in]
ax_1 x_2+cx_1 y_2, & {\rm if} \;\mathcal{R}=F^s.\\[0.05in]
 \end{array}\right.
 \end{eqnarray}
  For any $ax+cy\in\mathcal{R}$, where $x,y\in \mathbb{F}_q$, the Gray map $\phi$ from $\mathcal{R}$ to $\mathbb{F}_q^2$ is defined by
  $$\phi:\;\mathcal{R}\rightarrow\mathbb{F}_q^2,\;\;ax+cy\mapsto(y,x+y).$$
  The map $\phi$ is bijective, which can be extended naturally from $\mathcal{R}^n$ to $\mathbb{F}_{q}^{2n}$ of the form
    $$\phi:\;\mathcal{R}^n\rightarrow\mathbb{F}_q^{2n},\;\;ax+cy\mapsto(y,x+y),$$
    where $\mathcal{R}^n=a \mathbb{F}_{q}^n+c\mathbb{F}_{q}^n$ and $x,y\in\mathbb{F}_{q}^n$.

 Let $u,v$ be two vectors over $\mathbb{F}_q$, the Hamming weight of $u$ denoted by $wt_H(u)$ is the number of non-zero entries in $u$, and the Hamming distance of $u,v$ is $d_H(u,v)=wt_H(u-v)$.  A  code $\mathcal{C}$ of length $n$ over  $\mathcal{R}$ is an  $\mathcal{R}$-submodule of  $\mathcal{R}^n$. Let $r_1=ax_1+cy_1, r_2=ax_2+cy_2\in\mathcal{R}^n$, where $x_1,x_2,y_1,y_2\in\mathbb{F}_{q}^n$. The Lee weight of $r_1$ is the Hamming weight of its Gray image as $wt_L(r_1)=wt_H(\phi(r_1))=wt_H(y_1)+wt_H(x_1+y_1)$. The Lee distance between $r_1$ and $r_2$ is defined as $d_L(r_1,r_2)=wt_L(r_1-r_2)$. It is easy to check that the Gray map is an isometry from $(\mathcal{R}^n,d_L)$ to $(\mathbb{F}_{q}^{2n},d_H)$.   Let $A_{i}$ be the number of codewords $c\in\mathcal{C}$ with Lee weight $i$, $1\leq i\leq 2n$. Then the Lee weight enumerator of $\mathcal{C}$ is defined by $$1+A_{1}x+A_{2}x^{2}+\cdots+A_{2n}x^{2n}$$ and the sequence $(1,A_{1},\cdots,A_{2n})$ is called the Lee weight distribution of $\mathcal{C}$. An $n$-length code $\mathcal{C}$ over $\mathcal{R}$ is called a $t$-Lee-weight code if $|\{1\leq i\leq 2n: A_{i}\neq0\}|=t$. Moreover, we call the minimum non-zero Lee weight of  $\mathcal{C}$  the minimum Lee distance, which is denoted by $\min d_L$.
\subsection{ The Generic Construction of Linear Codes}

 Let $m$  be a positive integer. In 2007,  Ding and Niederreiter  \cite{CN}  initially proposed a generic construction of linear codes from subsets of $\mathbb{F}_{q^m}$  as follows. Let  $\Tr^{m}_{1}(\cdot)$ be the trace function from $\mathbb{F}_{q^m}$ to $\mathbb{F}_q$ and $D=\{d_{1},d_{2},\dots,d_{n}\}\subseteq\mathbb{F}_{q^m}\backslash\{0\}$. A linear code of length $n$ over $\mathbb{F}_q$ can be defined as
\begin{eqnarray*}
\mathcal{C}_D=\{c_D(v)=(\Tr^{m}_{1}(vx))_{x\in D}: v \in \mathbb{F}_{q^m}\}.
\end{eqnarray*}
The code $\mathcal{C}_D$ is called a trace code over $\mathbb{F}_q$ and the set $D$ is called the defining set of $\mathcal{C}_D$, which is a generalization of irreducible cyclic codes and the dimension of $\mathcal{C}_D$ is at most $m$. It is well known that many linear codes have been produced via selecting the defining set properly, see \cite{CC1,CNY,WLZ,JCC,JC,ZNCT} for example. Identifying $\mathbb{F}_{q^m}$ with $\mathbb{F}_q^m$, it has been proved that  the above trace code  is equivalent to
\begin{eqnarray}\label{D-Ding}
\mathcal{C}_D=\{c_D(v)=(\langle v, x\rangle_q)_{x\in D}: v \in \mathbb{F}_{q}^m\},
\end{eqnarray}
where $\langle x,y\rangle_q$ is the inner product of $x$ and $y$ over $\mathbb{F}_q$. In \cite{SLS,SGS}, the notation of the codes in \eqref{D-Ding} was extended from finite fields to finite rings as below.

%

Let   $\mathcal{R}^m$ be an extension of ring $\mathcal{R}$ of degree $m$ and  $L\subseteq\mathcal{R}^m$. Then  a code over $\mathcal{R}$ with a defining set $L$ is defined by
\begin{eqnarray}\label{Ring-Trace-Code}
\mathcal{C}_{L}=\{c_{L}(v)=(\langle v, x\rangle_q)_{x\in L}: v \in \mathcal{R}^m\}.
\end{eqnarray}
Some known codes over ring $\mathcal{R}$ have been constructed by choosing proper defining sets $L$. The researchers are referred to \cite{LS,SS,SS1,SS2,SL2} for example.

\subsection{Simplicial Complexes}
For $m\in\mathbb{N}$, we shall write $[m]$ to denote the set $\{1,2,\dots,m\}$.  For two subsets $A,B\subseteq[m]$, the set $\{x: x\in A\;\text{and}\;x\notin B\}$ is denoted by $A\backslash B$. The support  of a vector $\omega\in\mathbb{F}_{q}^m$
is defined by $\text{supp}(\omega)=\{i\in[m]:\,\omega_i\ne0\}$. Clearly, the Hamming weight $wt_H (\omega)$ of $\omega\in\mathbb{F}_{q}^m$ equals  $|\text{supp}(\omega)|$ which is the size of  $\text{supp}(\omega)$. Let $\nu\in\mathbb{F}_{q}^m$, one says that $\nu$ covers $\omega$ if $\text{supp}(\omega)\subseteq\text{supp}(\nu)$, which is written as $\omega\preceq \nu$. A subset $\Delta$ of $\mathbb{F}_q^m$ is called a simplicial complex if $\nu\in\Delta,\omega\in\mathbb{F}_{q}^m$ and  $\omega\preceq \nu$ imply $\omega\in\Delta$. An element $\nu\in\Delta$ is called a maximal element of $\Delta$ if $\nu$ is not properly covered in any other element of $\Delta$.

Now onwards, we will always write $F$ instead of $\text{supp}(F)$ whenever $F$ is a maximal element of simplicial complexes for simplify. According to the definition of simplicial complexes, for each $F \subseteq[m]$, the simplicial complex $\Delta_F$ of $\mathbb{F}_q^m$ generated by $F$ is defined to be
$$\Delta_F=\{\omega\in\mathbb{F}_q^m:\,\text{supp}(\omega)\subseteq F\}.$$
 Therefore, $\Delta_F$ is a vector subspace of $\mathbb{F}_q^m$ with dimension $|F|$. In particular, $\Delta_F=\{0\}$ if $F=\emptyset$.  A simplicial complex can have more than one maximal element. Let $\mathcal{F}= \{F_1,\,\dots,\,F_l \}$ be the family of maximal elements of $\Delta$, where $l$ is the number of maximal elements and $F_i\subseteq[m]$ is the maximal element of $\Delta$ for $i\in[l]$.  Then  $\Delta$ is  actually the union of $l$ vector spaces over $\mathbb{F}_q$ of dimensions  $|F_i|$.

\subsection{Our Construction and the Calculation of Lee Weight}
Based on the generic construction  of linear codes and simplicial complexes,  we propose our construction method by selecting proper rings and defining sets as below.

Hereafter, we always  assume that  $\mathcal{R}$ is an extension of the ring $E$ (or $F$)  of degree $s$ and  $L=aL_1+cL_2\subseteq\mathcal{R}^m$, where  $L_1,L_2\subseteq\mathbb{F}_{q}^{m}$ consisting of simplicial complexes and $q=p^s$. In this paper, we mainly study the codes over $\mathcal{R}$ defined by \eqref{Ring-Trace-Code} with the following five defining sets:
\begin{enumerate}
  \item $L=a\Delta_1+c\Delta_2$;
  \item $L=a\Delta_1^c+c\Delta_2$;
  \item  $L=a\Delta_1+c\Delta_2^c$;
  \item $L=a\Delta_1^c+c\Delta_2^c$, and
  \item $L=(a\Delta_1+c\Delta_2)^c$,
\end{enumerate}
    where $\Delta_1,\Delta_2$ are general simplicial complexes of $\mathbb{F}_q^m$ and $\Delta^c=\mathbb{F}_q^m\backslash\Delta$.  The length of  $\mathcal{C}_L$ is definitely $|L|=|L_1||L_2|$. In what follows, we give the Lee weight expression of codewords in $\mathcal{C}_L$.  Assume that $v=a\beta_1+c\beta_2\in\mathcal{R}^m$ and $x=at_1+ct_2\in L$, where $\beta_i\in\mathbb{F}_{q}^m$ and $t_i\in L_i$, $i=1,2$. Then
\begin{eqnarray}\label{vector-product}\langle v,x\rangle_q=\langle a\beta_1+c\beta_2, at_1+ct_2\rangle_q=\left\{\begin{array}{ll}
 a\langle\beta_1,t_1\rangle_q+c\langle\beta_2,t_1\rangle_q, & {\rm if}\; \mathcal{R}=E^s;\\[0.05in]
a\langle\beta_1,t_1\rangle_q+c\langle\beta_1,t_2\rangle_q, & {\rm if} \;\mathcal{R}=F^s\\[0.05in]
 \end{array}\right.\end{eqnarray} from \eqref{ring-multiplication}. Taking Gray map yields
 $$\phi((\langle v,x\rangle_q)_{x\in L})=\left\{\begin{array}{ll}
 (\langle\beta_2,t_1\rangle_q,\langle\beta_1+\beta_2,t_1\rangle_q)_{t_1\in L_1,t_2\in L_2}, & {\rm if}\; \mathcal{R}=E^s;\\[0.05in]
(\langle\beta_1,t_2\rangle_q,\langle\beta_1, t_1+t_2\rangle_q)_{t_1\in L_1,t_2\in L_2}, & {\rm if} \;\mathcal{R}=F^s.\\[0.05in]
 \end{array}\right.$$
 Recall that  $wt_L(r)=wt_H(\phi(r))$ for  $r\in\mathcal{R}^m$. Therefore, when $\mathcal{R}=E^s$, for any $v=a\beta_1+c\beta_2\in\mathcal{R}^m$, one has
\begin{eqnarray}\label{weight-for-E} wt_L(c_{L}(v))&=&wt_H( (\langle\beta_2,t_1\rangle_q)_{t_1\in L_1,t_2\in L_2})+wt_H((\langle\beta_1+\beta_2,t_1\rangle_q)_{t_1\in L_1,t_2\in L_2})\nonumber\\
&=&|L_1||L_2|-\frac{1}{q}\sum_{u\in\mathbb{F}_p^s}\sum_{t_1\in L_1}\sum_{t_2\in L_2}w_p^{\langle u,\langle\beta_2,t_1\rangle_q\rangle_p}+\nonumber\\
&&|L_1||L_2|-\frac{1}{q}\sum_{u\in\mathbb{F}_p^s}\sum_{t_1\in L_1}\sum_{t_2\in L_2}w_p^{\langle u,\langle\beta_1+\beta_2,t_1\rangle_q\rangle_p} \nonumber \\
&=&|L_2|\Big(2|L_1|-\frac{1}{q}\sum_{u\in\mathbb{F}_p^s}\sum_{t_1\in L_1}w_p^{\langle u,\langle\beta_2,t_1\rangle_q\rangle_p}-\frac{1}{q}\sum_{u\in\mathbb{F}_p^s}\sum_{t_1\in L_1}w_p^{\langle u,\langle\beta_1+\beta_2,t_1\rangle_q\rangle_p}\Big).
\end{eqnarray}
Similarly, we can deduce that when $\mathcal{R}=F^s$,
\begin{eqnarray}\label{weight-for-F} wt_L(c_{L}(v))=2|L_1||L_2|-\frac{1}{q}\Big(|L_1|\sum_{u\in\mathbb{F}_p^s}\sum_{t_2\in L_2}w_p^{\langle u,\langle\beta_1,t_2\rangle_q\rangle_p}+\sum_{u\in\mathbb{F}_p^s}\sum_{\substack{t_1\in L_1\\ t_2\in L_2}}w_p^{\langle u,\langle\beta_1,t_1+t_2\rangle_q\rangle_p}\Big).
\end{eqnarray}

\section{Useful Auxiliary Lemmas}
To calculate the lengths, sizes and the minimum Lee distance $\min d_L$ of these codes constructed in Section II-D, extensive preparatory work is required. In this section, we present several useful lemmas on exponential sums. We first give the following lemma which can be used to determine the size of the codes.

\begin{lem}\label{value-of-delta}Let $\Delta$ be a simplicial complex of $\mathbb{F}_q^m$ with the set of maximal elements $\mathcal{F}= \{F_1,\,\dots,\,F_l \}$. Then the size of $\Delta$ is
$$|\Delta|=\sum_{\emptyset\neq S\subseteq\mathcal{F}}(-1)^{|S|+1}q^{|\cap S|},$$
where $\cap S=\bigcap\limits_{F\in S}F$.
\end{lem}

\begin{IEEEproof} By using the principle of inclusion-exclusion, one has
\begin{eqnarray*}
  |\Delta|=|\Delta_{F_1}\cup\dots\cup\Delta_{F_l}|
  =\sum_{i\in[l]}|\Delta_{F_i}|-\sum_{1\leq i<j\leq l}|\Delta_{F_i}\cap\Delta_{F_j}|+\dots+(-1)^{l+1}|\cap_{i\in[l]}\Delta_{F_i}|.
\end{eqnarray*}
Then the result follows by the fact that $\Delta_{A}\cap\Delta_{B}=\Delta_{A\cap B}$ and $|\Delta_{A\cap B}|=q^{|A\cap B|}$ for any $A,B\subseteq[m]$.
\end{IEEEproof}

 Based on Lemma \ref{value-of-delta}, it can be seen that $|\Delta|<\sum_{i\in[l]}|\Delta_{F_i}|=\sum_{i\in[l]}q^{|F_i|}$. In what follows, we further discuss the relation between   $|\Delta|$ and $\sum_{i\in[l]}q^{|F_i|}$.

 \begin{lem}\label{delta-F1}Let $\Delta$ be a simplicial complex of $\mathbb{F}_q^m$ with the set of maximal elements $\mathcal{F}= \{F_1,\,\dots,\,F_l \}$. Let $F_i\backslash\cup_{j\in[l]\backslash\{i\}}F_j\neq\emptyset$ for any $i\in[l]$. Then
 \begin{eqnarray}\label{delta-F}
 2|\Delta|>\sum\nolimits_{i\in[l]}q^{|F_i|}.
 \end{eqnarray}
 \end{lem}

 \begin{IEEEproof} When $l=1$, it is clear that $2|\Delta|=2q^{|F_1|}>q^{|F_1|}$. When $l=2$, by Lemma  \ref{value-of-delta}, one has
 $$|\Delta|=|\Delta_{F_1}|+|\Delta_{F_2}|-|\Delta_{F_1}\cap\Delta_{F_2}|=q^{|F_1|}+q^{|F_2|}-q^{|F_1\cap F_2|},$$
 which induces that $2|\Delta|-(q^{|F_1|}+q^{|F_2|})=q^{|F_1|}+q^{|F_2|}-2q^{|F_1\cap F_2|}>0$ due to $|F_1\cap F_2|<\min\{|F_1|,|F_2|\}$. Therefore, \eqref{delta-F} is satisfied for $l=2$. By induction, we assume that the inequality holds for $l=t$ ($t\geq 1$). Then when $l=t+1$, one has
 \begin{eqnarray*}
 2|\Delta|&=&2\left(|\Delta_{F_1}\cup\dots\cup\Delta_{F_t}|+|\Delta_{F_{t+1}}|-|(\Delta_{F_1}\cup\dots\cup\Delta_{F_t})\cap\Delta_{F_{t+1}}|\right)\\
 &>&\sum\nolimits_{i\in[t]}q^{|F_i|}+2q^{|F_{t+1}|}-2|(\Delta_{F_1}\cup\dots\cup\Delta_{F_t})\cap\Delta_{F_{t+1}}|\\
 &=&\sum\nolimits_{i\in[t+1]}q^{|F_i|}+q^{|F_{t+1}|}-2|(\Delta_{F_1}\cup\dots\cup\Delta_{F_t})\cap\Delta_{F_{t+1}}|.
 \end{eqnarray*}
 Observe that $\Delta_{F_1}\cup\dots\cup\Delta_{F_t}\subseteq\Delta_{\cup_{i\in[t]}F_i}$. Hence,
 $$|(\Delta_{F_1}\cup\dots\cup\Delta_{F_t})\cap\Delta_{F_{t+1}}|<|\Delta_{\cup_{i\in[t]}F_i}\cap\Delta_{F_{t+1}}|=|\Delta_{(\cup_{i\in[t]}F_i)\cap F_{t+1}}|\leq q^{|F_{t+1}|-1}$$
 since  $F_{t+1}\backslash\cup_{j\in[t]}F_j\neq\emptyset$ . Further, we can derive that
 $$2|\Delta|>\sum\nolimits_{i\in[t+1]}q^{|F_i|}+q^{|F_{t+1}|}-2q^{|F_{t+1}|-1}\geq \sum\nolimits_{i\in[t+1]}q^{|F_i|}.$$
 This completes the proof.
 \end{IEEEproof}

\begin{lem}\label{cap-cup-bot}Let $m$, $l$ be positive integers and $F_i\subseteq[m]$ for $ i\in[ l]$.
\begin{enumerate}
    \item [(1)]  For any $i,j\in[l]$, if $\Delta_{F_i}\subseteq\Delta_{F_j}$, then $\Delta_{F_j}^{\bot}\subseteq\Delta_{F_i}^{\bot}$.

 \item [(2)]  $\cap_{i\in[l]}\Delta_{F_i}^{\bot}=\Delta_{\cup_{i\in[l]}F_i}^{\bot}$.
 \end{enumerate}
\end{lem}

\begin{IEEEproof}
(1) Assume $x\in\Delta_{F_j}^{\bot}$. For any $y\in\Delta_{F_i}$, we have $y\in\Delta_{F_j}$ due to $\Delta_{F_i}\subseteq\Delta_{F_j}$. Hence, one has $\langle x,y\rangle_q=0$ according to the definition of the dual space. This gives $x\in\Delta_{F_i}^{\bot}$, which implies that $\Delta_{F_j}^{\bot}\subseteq\Delta_{F_i}^{\bot}$.

(2)  Since $\Delta_{F_i}^{\bot}=\Delta_{[m]\backslash F_i}$ and $\Delta_{F_i}\cap \Delta_{F_j}=\Delta_{F_i\cap F_j}$ for any $i,j\in[l]$,  we then have  $\cap_{i\in[l]}\Delta_{F_i}^{\bot}=\cap_{i\in[l]}\Delta_{[m]\backslash F_i}=\Delta_{\cap_{i\in[l]} ([m]\backslash F_i)}=\Delta_{[m]\backslash (\cup_{i\in[l]}F_i)}=\Delta_{\cup_{i\in[l]}F_i}^{\bot}$. This completes the proof.
\end{IEEEproof}

One can see from \eqref{weight-for-E} and \eqref{weight-for-F} that in order to determine the minimum Lee distance of $\mathcal{C}_L$,  it is necessary to discuss the possible values of $wt_L(\mathcal{C}_L(v))$ when $v$ runs through $\mathcal{R}^m$. Firstly, we give more explicit expressions of  the exponential sums involved in  \eqref{weight-for-E} and \eqref{weight-for-F} by virtue of the duality of vector spaces and the orthogonality of exponential sums, see Lemmas  \ref{vector-space-exponent-sum}-\ref{two-simplicial-comp-exponent-sum}.

Let    $m,s$ be  positive integers,  $q=p^s$ and $H$ be an $\upsilon$-dimensional $\mathbb{F}_q$-subspace of $\mathbb{F}_{q}^m$. The dual of $H$ is
$$H^{\bot}=\{x\in\mathbb{F}_q^m: \langle x,y\rangle_q=0\;\text{for any}\; y\in H\}.$$
It is clear that  $H^{\bot}$ is an $(m-\upsilon)$-dimensional $\mathbb{F}_q$-subspace of $\mathbb{F}_{q}^m$. Suppose  that  $\Delta_F$ is a simplicial complex of $\mathbb{F}_q^m$ with one  maximal element $F\subseteq[m]$, then  one can immediately  get that  $\Delta_F^{\bot}=\Delta_{[m]\backslash F}$ since $\Delta_F$ is an $|F|$-dimensional $\mathbb{F}_q$-subspace of $\mathbb{F}_{q}^m$. Let $u\in\mathbb{F}_{p}^s$ and $x,y\in\mathbb{F}_q^m$. The operator $\langle u,\langle y,x\rangle_q\rangle_p$ is reasonable when identifying  the element $\langle y,x\rangle_q\in\mathbb{F}_q$ with  the element in $\mathbb{F}_p^s$ due to $\mathbb{F}_q\cong\mathbb{F}_p^s$. Then we have the following lemma.

\begin{lem}\label{vector-space-exponent-sum}Let    $m,s$ be  positive integers and  $q=p^s$. Let $H$ be an $\upsilon$-dimensional $\mathbb{F}_q$-subspace of $\mathbb{F}_{q}^m$ and   $ u\in\mathbb{F}_{p}^{s}\backslash\{0\}$,  one has
$$\sum_{x\in H}w_p^{\langle u,\langle y,x\rangle_q\rangle_p}=\left\{\begin{array}{ll}
 q^\upsilon, & {\rm if}\;y\in H^{\bot};\\[0.05in]
0, & {\rm otherwise}.\\[0.05in]
 \end{array}\right.$$
\end{lem}

\begin{IEEEproof} If $y\in H^{\bot}$, then $\langle u,\langle y,x\rangle_q\rangle_p=\langle u,0\rangle_p=0$ for any $x\in H$, which induces that $\sum_{x\in H}w_p^{\langle u,\langle y,x\rangle_q\rangle_p}=q^\upsilon$. If $y\notin H^{\bot}$, then there is some $x_0\in H$ such that $\langle y,x_0\rangle_q=b\in\mathbb{F}_q^*$. Note that for any $t\in\mathbb{F}_q^*$, one has $\langle y,tx_0\rangle_q=bt\in\mathbb{F}_q^*$. Therefore, when $t$ runs over $\mathbb{F}_q$, so does  $\langle y,tx_0\rangle_q$. It indicates that there must be some $t_0\in\mathbb{F}_q$ satisfies $\langle u,\langle y,t_0x_0\rangle_q\rangle_p\ne0$. Since $H=t_0x_0+ H $,
$$\sum_{x\in H}w_p^{\langle u,\langle y,x\rangle_q\rangle_p}=\sum_{x\in H}w_p^{\langle u,\langle y,x+t_0x_0\rangle_q\rangle_p}=w_p^{\langle u,\langle y,t_0x_0\rangle_q\rangle_p}\sum_{x\in H}w_p^{\langle u,\langle y,x\rangle_q\rangle_p}.$$
The result follows by $w_p^{\langle u,\langle y,t_0x_0\rangle_q\rangle_p}\ne 1$ due to $\langle u,\langle y,t_0x_0\rangle_q\rangle_p\ne0$. This completes the proof.
\end{IEEEproof}

Denote by  $\alpha(\beta| \Delta_A)$   a $\{0,1\}$-valued function of $\mathbb{F}_{q}^m$ and $\delta_{i,j}$  the kronecker delta function, namely,
\begin{eqnarray}\label{alpha-delta}
\alpha(\beta| \Delta_A)=\left\{\begin{array}{ll}
 1, & {\rm if}\;\beta\in \Delta_A^{\bot};\\[0.05in]
0, & {\rm otherwise}\\[0.05in]
 \end{array}\right.
 \;{\rm and}\;\;
\delta_{i,j}=\left\{\begin{array}{ll}
 1, & {\rm if}\;j=i;\\[0.05in]
0, & {\rm otherwise.}\\[0.05in]
 \end{array}\right.
\end{eqnarray}
Then we have the following results.
\begin{lem}\label{a-simplicial-comp-exponent-sum}Let $q=p^s$ and $\Delta$ be a simplicial complex of $\mathbb{F}_q^m$ with $\mathcal{F}=\{F_1,\dots,F_l\}$ the set of maximal elements of $\Delta$. For any $\beta\in\mathbb{F}_q^m$, we have
$$ \sum_{u\in\mathbb{F}_p^s}\sum_{t\in\Delta}w_p^{\langle u,\langle\beta,t\rangle_q\rangle_p}=\sum_{\emptyset\neq S\subseteq\mathcal{F}}(-1)^{|S|+1}q^{|\cap S|}(1+(q-1)\alpha(\beta|\Delta_{\cap S}))$$
and
$$ \sum_{u\in\mathbb{F}_p^s}\sum_{t\in\Delta^c}w_p^{\langle u,\langle\beta,t\rangle_q\rangle_p}=q^m+(q-1)q^m\delta_{0,\beta}-\sum_{u\in\mathbb{F}_p^s}\sum_{t\in\Delta}w_p^{\langle u,\langle\beta,t\rangle_q\rangle_p},$$
where $\cap S=\bigcap\limits_{F\in S}F$, $\alpha(\beta| \Delta_A)$ and  $\delta_{i,j}$ are defined in \eqref{alpha-delta}.
\end{lem}
\begin{IEEEproof}For any $F\subseteq [m]$, one can obtain that
$$ \sum_{u\in\mathbb{F}_p^s}\sum_{t\in\Delta_F}w_p^{\langle u,\langle\beta,t\rangle_q\rangle_p}=q^{|F|}+ \sum_{0\ne u\in\mathbb{F}_p^s}\sum_{t\in\Delta_F}w_p^{\langle u,\langle\beta,t\rangle_q\rangle_p}=q^{|F|}+(q-1)q^{|F|}\alpha(\beta|\Delta_{F}) $$
due to Lemma \ref{vector-space-exponent-sum}. Therefore, using  the principle of inclusion-exclusion gives that
\begin{align*}
 \sum_{u\in\mathbb{F}_p^s}\sum_{t\in\Delta}w_p^{\langle u,\langle\beta,t\rangle_q\rangle_p}=&\sum_{i\in[l]}\Big( \sum_{u\in\mathbb{F}_p^s}\sum_{t\in\Delta_{F_i}}w_p^{\langle u,\langle\beta,t\rangle_q\rangle_p}\Big)-
\sum_{1\leq i<j\leq l}\Big( \sum_{u\in\mathbb{F}_p^s}\sum_{t\in\Delta_{F_i}\cap\Delta_{F_j}}w_p^{\langle u,\langle\beta,t\rangle_q\rangle_p}\Big)\\&+\dots+(-1)^{l+1}\Big( \sum_{u\in\mathbb{F}_p^s}\sum_{t\in\cap_{i\in[l]}\Delta_{F_i}}w_p^{\langle u,\langle\beta,t\rangle_q\rangle_p}\Big)\\
=&\sum_{\emptyset\neq S\subseteq\mathcal{F}}(-1)^{|S|+1}(q^{|\cap S|}+(q-1)q^{|\cap S|}\alpha(\beta|\Delta_{\cap S})).
\end{align*}
The second assertion follows from the fact that $\Delta^c=\mathbb{F}_q^m\backslash\Delta$ and $ \sum_{u\in\mathbb{F}_p^s}\sum_{t\in\mathbb{F}_q^m}w_p^{\langle u,\langle\beta,t\rangle_q\rangle_p}=q^m+(q-1)q^m\delta_{0,\beta}$. This completes the proof.
\end{IEEEproof}

\begin{lem}\label{two-simplicial-comp-exponent-sum}Let $q=p^s$ and  $\Delta_1$, $\Delta_2$  be two simplicial complexes of $\mathbb{F}_q^m$ with $\mathcal{F}=\{F_1,\dots,F_l\}$ and $\mathcal{H}=\{H_1,\dots,H_k\}$ being the sets of maximal elements of $\Delta_1$ and $\Delta_2$, respectively. For any $\beta\in\mathbb{F}_q^m$, we have
\begin{align*} \sum_{u\in\mathbb{F}_p^s}\sum_{\substack{t_1\in\Delta_1 \\ t_2\in\Delta_2}}w_p^{\langle u,\langle\beta,t_1+t_2\rangle_q\rangle_p}=
\sum_{\substack{\emptyset\neq S_1\subseteq\mathcal{F}\\ \emptyset\neq S_2\subseteq\mathcal{H}}}(-1)^{|S_1|+|S_2|}q^{|\cap S_1|+|\cap S_2|}(1+(q-1)\alpha(\beta|\Delta_{\cap S_1})\alpha(\beta|\Delta_{\cap S_2}))
\end{align*}
and
\begin{align*} \sum_{u\in\mathbb{F}_p^s}\sum_{\substack{t_1\in\Delta_1 \\ t_2\in\Delta_2^c}}w_p^{\langle u,\langle\beta,t_1+t_2\rangle_q\rangle_p}=q^m|\Delta_1|(1+(q-1)\delta_{0,\beta})- \sum_{u\in\mathbb{F}_p^s}\sum_{\substack{t_1\in\Delta_1 \\ t_2\in\Delta_2}}w_p^{\langle u,\langle\beta,t_1+t_2\rangle_q\rangle_p},
\end{align*}
where $\cap S=\bigcap\limits_{F\in S}F$,  $\alpha(\beta| \Delta_A)$ and  $\delta_{i,j}$ are defined in \eqref{alpha-delta}.
\end{lem}
\begin{IEEEproof}Similarly as the proof of Lemma \ref{a-simplicial-comp-exponent-sum}, one has
\begin{align*}
&\sum_{u\in\mathbb{F}_p^s}\sum_{t_1\in\Delta_1}\sum_{t_2\in\Delta_2}w_p^{\langle u,\langle\beta,t_1+t_2\rangle_q\rangle_p}\\
=&\sum_{u\in\mathbb{F}_p^s}\Big( \sum_{\emptyset\neq S_1\subseteq\mathcal{F}}(-1)^{|S_1|+1}\sum_{t_1\in \Delta_{\cap S_1}} w_p^{\langle u,\langle\beta,t_1\rangle_q\rangle_p}\Big)\Big( \sum_{\emptyset\neq S_2\subseteq\mathcal{H}}(-1)^{|S_2|+1}\sum_{t_2\in \Delta_{\cap S_2}} w_p^{\langle u,\langle\beta,t_2\rangle_q\rangle_p}\Big)\\
=&\sum_{\substack{\emptyset\neq S_1\subseteq\mathcal{F} \\ \emptyset\neq S_2\subseteq\mathcal{H}}}(-1)^{|S_1|+|S_2|}q^{|\cap S_1|+|\cap S_2|}
+\sum_{0\ne u\in\mathbb{F}_p^s}\sum_{\substack{\emptyset\neq S_1\subseteq\mathcal{F} \\ \emptyset\neq S_2\subseteq\mathcal{H}}}(-1)^{|S_1|+|S_2|}q^{|\cap S_1|+|\cap S_2|}\alpha(\beta|\Delta_{\cap S_1})\alpha(\beta|\Delta_{\cap S_2})\\
=&\sum_{\substack{\emptyset\neq S_1\subseteq\mathcal{F} \\ \emptyset\neq S_2\subseteq\mathcal{H}}}(-1)^{|S_1|+|S_2|}q^{|\cap S_1|+|\cap S_2|}(1+(q-1)\alpha(\beta|\Delta_{\cap S_1})\alpha(\beta|\Delta_{\cap S_2})),
\end{align*}
where the second equality holds due to Lemma \ref{vector-space-exponent-sum}. This completes the proof.
\end{IEEEproof}

According to Lemmas \ref{cap-cup-bot}-\ref{two-simplicial-comp-exponent-sum}, several important lemmas, which is useful to proof the main results of this paper can be obtained   as below.

\begin{lem}\label{linear-weight}Let $m$, $l$, $s$ be positive integers, $q=p^s$ and $\Delta$ be a simplicial complex of $\mathbb{F}_q^m$ with $\mathcal{F}=\{F_1,\dots,F_l\}$ the set of maximal elements of $\Delta$. Define
$$\mathcal{A}_{\Delta,\beta}=|\Delta|-\frac{1}{q}\sum_{u\in\mathbb{F}_p^s}\sum_{t\in \Delta}w_p^{\langle u,\langle\beta,t\rangle_q\rangle_p}
$$
and
$$f_{\beta}(S)=(-1)^{|S|+1}q^{|\cap S|-1}(q-1)\left(1-\alpha(\beta|\Delta_{\cap S})\right),$$
where $\beta\in\mathbb{F}_q^m$, $\cap S=\bigcap_{F\in S}F$ and $\alpha(\beta| \Delta_A)$ is  shown in \eqref{alpha-delta}. Then
$\mathcal{A}_{\Delta,\beta}=\sum_{\emptyset\neq S\subseteq\mathcal{F}}f_{\beta}(S)$ and
\begin{enumerate}
    \item [(1)] $\mathcal{A}_{\Delta,\beta}=0$ if and only if  $\beta\in\cap_{i\in[l]}\Delta_{F_i}^{\bot}$, that is $\beta\in \Delta_{\cup_{i\in[l]}F_i}^{\bot}$;

  \item [(2)]   If $\beta\notin\Delta_{F_i}^{\bot}$ for some $i\in[l]$, then $\mathcal{A}_{\Delta,\beta}\geq(q-1)q^{|F_i|-1}$. Moreover, the equality holds if  $\beta\notin\Delta_{F_i}^{\bot}$ and $\beta\in\Delta_{\cup_{j\in[l]\backslash\{i\}}F_j}^{\bot}$;

  \item [(3)]   $\mathcal{A}_{\Delta,\beta}\leq(q-1)\sum_{i\in[l]}q^{|F_i|-1}$, and the equality holds if and only if $\beta\notin\cup_{i\in[l]}\Delta_{F_i}^{\bot}$ and $\beta\in \cap_{1\leq i<j\leq l}\Delta_{F_i\cap F_j}^{\bot}$.
\end{enumerate}
\end{lem}

\begin{IEEEproof}  It should be mentioned that $\mathcal{A}_{\Delta,\beta}$ is actually the Hamming weight of $c_{D}(\beta)$ in $\mathcal{C}_D$ defined by \eqref{D-Ding}, where $D=\Delta$.  When $l=1$, the result can be easily obtained and thus, we mainly focus on the case that $l\geq 2$. From Lemmas \ref{value-of-delta} and \ref{a-simplicial-comp-exponent-sum}, one has   that
\begin{eqnarray}\label{weight-value}
\mathcal{A}_{\Delta,\beta}&=&\sum_{\emptyset\neq S\subseteq\mathcal{F}}(-1)^{|S|+1}q^{|\cap S|}
\Big(1-\frac{1}{q}(1+(q-1)\alpha(\beta|\Delta_{\cap S}))\Big)\nonumber\\
&=&\sum_{\emptyset\neq S\subseteq\mathcal{F}}(-1)^{|S|+1}q^{|\cap S|-1}(q-1)\left(1-\alpha(\beta|\Delta_{\cap S})\right)\nonumber\\
&=&\sum_{\emptyset\neq S\subseteq\mathcal{F}}f_{\beta}(S).
\end{eqnarray}
Denote by
  $\mathcal{F}_i=\{F_1\cap F_i, F_2\cap F_i,\dots,F_{i-1}\cap F_i\}$ for $2\leq i\leq l$ and $\mathcal{F}_1=\emptyset$ for convenience.  With the notation above, we claim that
  \begin{eqnarray}\label{f_u(s)}
  \sum_{\emptyset\neq S\subseteq\mathcal{F}} f_{\beta}(S)=\sum_{i\in[l]}\sum_{\emptyset\neq S\subseteq\{F_i\}} f_{\beta}(S)-\sum_{i\in[l]\backslash\{1\}}\sum_{\emptyset\neq S\subseteq\mathcal{F}_i} f_{\beta}(S).
    \end{eqnarray}
    When $l=2$, it is clear  that
    $$\sum_{\emptyset\neq S\subseteq\{F_1,F_2\}} f_{\beta}(S)=\sum_{\emptyset\neq S\subseteq\{F_1\}} f_{\beta}(S)+\sum_{\emptyset\neq S\subseteq\{F_2\}} f_{\beta}(S)-\sum_{\emptyset\neq S\subseteq\mathcal{F}_2} f_{\beta}(S),$$
    which is consistent with \eqref{f_u(s)}. By induction, we assume that the equality holds for  $l\geq2$. Then one has
    \begin{eqnarray*}\sum_{\emptyset\neq S\subseteq\{F_1,\dots,F_{l+1}\}} f_{\beta}(S)&=&\sum_{\emptyset\neq S\subseteq\{F_1,\dots,F_{l}\}} f_{\beta}(S)+\sum_{\emptyset\neq S\subseteq\{F_{l+1}\}} f_{\beta}(S)-\sum_{\emptyset\neq S\subseteq\mathcal{F}_{l+1}} f_{\beta}(S)\\
    &=&\sum_{i\in[l+1]}\sum_{\emptyset\neq S\subseteq\{F_i\}} f_{\beta}(S)-\sum_{i\in[l+1]\backslash\{1\}}\sum_{\emptyset\neq S\subseteq\mathcal{F}_i} f_{\beta}(S),
    \end{eqnarray*}
 which indicates that \eqref{f_u(s)} holds for any $l\geq 2$. Further, \eqref{weight-value} together with \eqref{f_u(s)} gives that
 \begin{eqnarray}\label{weight-value-f_u(s)}
 \mathcal{A}_{\Delta,\beta}=\sum_{\emptyset\neq S\subseteq\mathcal{F}}f_{\beta}(S)=
\sum_{i\in[l]}\sum_{\emptyset\neq S\subseteq\{F_i\}} f_{\beta}(S)-\sum_{i\in[l]\backslash\{1\}}\sum_{\emptyset\neq S\subseteq\mathcal{F}_i} f_{\beta}(S)
  \end{eqnarray}
  Now we shall complete the proof by virtue of \eqref{weight-value-f_u(s)}.

  (1)  Observe that for any  $J_1\subseteq J_2\subseteq[l]$, one has $$\Delta_{\cap_{j\in J_1}F_j}^{\bot}\subseteq\Delta_{\cap_{j\in J_2}F_j}^{\bot}\subseteq\Delta_{\cap_{j\in[l]}F_j}^{\bot}$$
due to $\Delta_{\cap_{j\in[l]}F_j}\subseteq\Delta_{\cap_{j\in J_2}F_j}\subseteq\Delta_{\cap_{j\in J_1}F_j}$ and  Lemma \ref{cap-cup-bot}(1). Therefore, one can get that if $\beta\in\cap_{i\in[l]}\Delta_{F_i}^{\bot}$, then $\beta\in\Delta_{\cap S}^{\bot}$  for any $\emptyset\neq S\subseteq\mathcal{F}$, which suggests that
  $1-\alpha(\beta|\Delta_{\cap S})=0$ for any $\emptyset\neq S\subseteq\mathcal{F}$. Therefore, one has $\mathcal{A}_{\Delta,\beta}=0$ due to \eqref{weight-value}.  On the other hand, suppose
 $\beta\notin\cap_{i\in[l]}\Delta_{F_i}^{\bot}$, then there must exist some $i\in[l]$ such that $\beta\notin\Delta_{F_i}^{\bot}$. Without loss of generality, we assume that $\beta\notin \Delta_{F_1}^{\bot}$. According to \eqref{weight-value-f_u(s)}, one has
 \begin{eqnarray}\label{A}
 \mathcal{A}_{\Delta,\beta}=\sum_{\emptyset\neq S\subseteq\{F_1\}} f_{\beta}(S)+\sum_{i\in[l]\backslash\{1\}}\Big(\sum_{\emptyset\neq S\subseteq\{F_i\}} f_{\beta}(S)-\sum_{\emptyset\neq S\subseteq\mathcal{F}_i} f_{\beta}(S)\Big).
  \end{eqnarray}
Note that for any $ i\in[ l]\backslash\{1\}$, $\sum_{\emptyset\neq S\subseteq\{F_i\}} f_{\beta}(S)-\sum_{\emptyset\neq S\subseteq\mathcal{F}_i} f_{\beta}(S)$ is the Hamming weight of $c_D(v)$ with $D=\Delta_{F_i}\backslash\Delta_{\mathcal{F}_i}$ by comparing with $\mathcal{A}_{\Delta,\beta}$. Therefore,
\begin{eqnarray*}\mathcal{A}_{\Delta,\beta}\geq\sum_{\emptyset\neq S\subseteq\{F_1\}} f_{\beta}(S)=q^{|F_1|-1}(q-1)\left(1-\alpha(\beta|\Delta_{F_1})\right)
=q^{|F_1|-1}(q-1)>0
\end{eqnarray*} since   $\beta\notin\Delta_{F_1}^{\bot}$. Thus, we conclude that $\mathcal{A}_{\Delta,\beta}=0$ if and only if  $\beta\in\cap_{i\in[l]}\Delta_{F_i}^{\bot}$, which is equivalent to $\beta\in \Delta_{\cup_{i\in[l]}F_i}^{\bot}$ due to Lemma \ref{cap-cup-bot}(2).

(2) From the proof above, one has that $\mathcal{A}_{\Delta,\beta}\geq(q-1)q^{|F_1|-1}$ if   $\beta\notin\Delta_{F_1}^{\bot}$. When $\beta\notin\Delta_{F_i}^{\bot}$  for some $i\in[l]\backslash\{1\}$, with a similar process and  reversing the positions of $F_1$ and $F_i$, one can also obtain
  \begin{eqnarray}\label{compare-weight}
  \mathcal{A}_{\Delta,\beta}\geq\sum_{\emptyset\neq S\subseteq\{F_i\}} f_{\beta}(S)=q^{|F_i|-1}(q-1).
  \end{eqnarray}
  Therefore, we only consider the case  $i=1$ for convenience in the following.

  Note that when $\beta\notin\Delta_{F_1}^{\bot}$, we have
  $\sum_{\emptyset\neq S\subseteq\{F_1\}} f_{\beta}(S)=(q-1)q^{|F_1|-1}$ due to $\alpha(\beta|\Delta_{F_1})=0$. Besides, if    $\beta\in\Delta_{\cup_{j\in[l]\backslash\{1\}}F_j}^{\bot}$, one then has $\beta\in\Delta_{F_j}^{\bot}$ for any $j\in[l]\backslash\{1\}$ and $\beta\in\Delta_{ F_i\cap F_j}^{\bot}$ for any $1\leq i<j\leq l$ due to $F_j\subseteq\cup_{i\in[l]\backslash\{1\}}F_i$ for any $j\in[l]\backslash\{1\}$, $ F_i\cap F_j\subseteq\cup_{j\in[l]\backslash\{1\}}F_j$ and Lemma \ref{cap-cup-bot}(1). Therefore, when $\beta\in\Delta_{\cup_{j\in[l]\backslash\{1\}}F_j}^{\bot}$, one can obtain  from Lemma \ref{linear-weight}(1)  that
   $$\sum_{\emptyset\neq S\subseteq\{F_i\}} f_{\beta}(S)=\sum_{\emptyset\neq S\subseteq\mathcal{F}_i} f_{\beta}(S)=0$$
    for any $i\in[l]\backslash\{1\}$.
     Further, by \eqref{A}, we have
    $$\mathcal{A}_{\Delta,\beta}=\sum_{\emptyset\ne S\subseteq\{F_1\}}f_\beta(S)=(q-1)q^{|F_1|-1}$$
    when  $\beta\notin\Delta_{F_1}^{\bot}$ and $\beta\in\Delta_{\cup_{j\in[l]\backslash\{1\}}F_j}^{\bot}$. This proves Lemma \ref{linear-weight}(2).

  (3) Again by \eqref{weight-value-f_u(s)}, one has
 \begin{eqnarray*}
 \mathcal{A}_{\Delta,\beta}&=&\sum_{i\in[l]}\sum_{\emptyset\neq S\subseteq\{F_i\}} f_{\beta}(S)-\sum_{i\in[l]\backslash\{1\}}\sum_{\emptyset\neq S\subseteq\mathcal{F}_i} f_{\beta}(S)\\
 &=&\sum_{i\in[l]}\sum_{\emptyset\neq S\subseteq\{F_i\}} f_{\beta}(S)-\sum_{i\in[l]\backslash\{1\}} wt_H(c_{\Delta_{\mathcal{F}_i}}(\beta)).
  \end{eqnarray*}
 Thus,  $$ \mathcal{A}_{\Delta,\beta}\leq\sum_{i\in[l]}\sum_{\emptyset\neq S\subseteq\{F_i\}} f_{\beta}(S)\leq\sum_{i\in[l]}(q-1)q^{|F_i|-1}.$$ The forgoing equality holds if
and only if $\sum_{\emptyset\neq S\subseteq\{F_i\}} f_{\beta}(S)=(q-1)q^{|F_i|-1}$ for any $i\in[l]$ and $wt_H(c_{\Delta_{\mathcal{F}_i}}(\beta))=0$ for any   $i\in[l]\backslash\{1\}$, which is equivalent to
\begin{eqnarray*}
\beta\notin\Delta_{F_i}^{\bot}\; \text{for any}\; i\in[l]\; \text{and}\; \beta\in\cap_{1\leq i<j\leq l}\Delta_{F_i\cap F_j}^{\bot}
\end{eqnarray*}
 according to the notation of $ f_{\beta}(S)$ and  Lemma \ref{linear-weight}(1).  This proves Lemma \ref{linear-weight}(3).
\end{IEEEproof}

\begin{lem}\label{two-linear-weight}Let $m$, $s$, $l$, $k$ be positive integers, $q=p^s$ and $\beta\in\mathbb{F}_q^m$. Let $\Delta_1$ and $\Delta_2$ be two simplicial complexes of $\mathbb{F}_q^m$ with  the sets of maximal elements $\mathcal{F}=\{F_1,\dots,F_l\}$ and $\mathcal{H}=\{H_1,\dots,H_k\}$, respectively.  Let
$ F_i\backslash\cup_{\ell\in[l]\backslash\{i\}}F_\ell\neq\emptyset$ and $ H_j\backslash\cup_{t\in[k]\backslash\{j\}}F_t\neq\emptyset$ for any $i\in[l]$,  $j\in[k]$.  Define
$$\tilde{\mathcal{A}}_\beta=|\Delta_1|| \Delta_2|-\frac{1}{q}\sum_{u\in\mathbb{F}_p^s}\sum_{t_1\in \Delta_1}\sum_{t_2\in \Delta_2}w_p^{\langle u,\langle\beta,t_1+t_2\rangle_q\rangle_p}
$$
and
$$\mathcal{T}=|\Delta_1|\mathcal{A}_{\Delta_2,\beta}+\tilde{\mathcal{A}}_\beta,$$
where $\mathcal{A}_{\Delta,\beta}$ is defined in Lemma \ref{linear-weight}. Then we have
\begin{enumerate}
    \item [(1)]  $\mathcal{T}=0$ if and only if  $\beta\in\cap_{i\in[l]}\Delta_{F_i}^{\bot}$ and  $\beta\in\cap_{j\in[k]}\Delta_{H_j}^{\bot}$, that is $\beta\in \Delta_{(\cup_{i\in[l]}F_i)\cup(\cup_{j\in[k]}H_j)}^{\bot}$;

  \item [(2)] The minimum non-zero value of $\mathcal{T}$ satisfies $$\mathcal{T}\geq (q-1) \mathop{\min}_{i\in [l],j\in[k]}\{|\Delta_2|q^{|F_i|-1},2|\Delta_1|q^{|H_j|-1}\}.$$
   Moreover, $\mathcal{T}= (q-1)|\Delta_2|q^{|F_i|-1}$  for some $i\in[l]$  if
$\beta\notin\Delta_{F_i}^{\bot}$ and  $\beta\in\Delta_{(\cup_{\ell\in[l]\backslash\{i\}}F_\ell)\cup(\cup_{j\in[k]}H_j)}^{\bot}$;  $\mathcal{T}= 2(q-1)|\Delta_1|q^{|H_j|-1}$  for some $j\in[k]$  if  $\beta\notin\Delta_{H_j}^{\bot}$ and $\beta\in\Delta_{(\cup_{t\in[k]\backslash\{j\}}H_t)\cup(\cup_{i\in[l]}F_i)}^{\bot}$;

  \item [(3)]  The maximum non-zero value of $\mathcal{T}$ satisfies $$\mathcal{T}\leq 2(q-1)|\Delta_1|\sum\nolimits_{j\in[k]}q^{|H_j|-1}.$$ The equality holds if and only if $\beta\notin\cup_{j\in[k]}\Delta_{H_j}^{\bot}$ and $\beta\in\Delta_{( \cup_{1\leq i<j\leq k}H_i\cap H_j)\cup(\cup_{i\in[l]}F_i)}^{\bot}$.
\end{enumerate}
\end{lem}


\begin{IEEEproof}  From Lemmas \ref{value-of-delta} and \ref{two-simplicial-comp-exponent-sum}, one has   that
\begin{eqnarray}\label{weight-value-2}
\tilde{\mathcal{A}}_\beta&=&\sum_{\substack{\emptyset\neq S_1\subseteq\mathcal{F}\\ \emptyset\neq S_2\subseteq\mathcal{H}}}(-1)^{|S_1|+|S_2|}q^{|\cap S_1|+|\cap S_2|}
\Big(1-\frac{1}{q}(1+(q-1)\alpha(\beta|\Delta_{\cap S_1})\alpha(\beta|\Delta_{\cap S_2}))\Big)\nonumber\\
&=&\sum_{\substack{\emptyset\neq S_1\subseteq\mathcal{F}\\ \emptyset\neq S_2\subseteq\mathcal{H}}}(-1)^{|S_1|+|S_2|}q^{|\cap S_1|+|\cap S_2|-1}(q-1)\left(1-\alpha(\beta|\Delta_{\cap S_1})\alpha(\beta|\Delta_{\cap S_2})\right)\nonumber\\
&=&\frac{q-1}{q}\Big(|\Delta_1||\Delta_2|-\sum_{\emptyset\neq S_1\subseteq\mathcal{F}}(-1)^{|S_1|+1}q^{|\cap S_1|}\alpha(\beta|\Delta_{\cap S_1})\sum_{\emptyset\neq S_2\subseteq\mathcal{H}}(-1)^{|S_2|+1}q^{|\cap S_2|}\alpha(\beta|\Delta_{\cap S_2})\Big).
\end{eqnarray}
Note that
$$\sum_{\emptyset\neq S_1\subseteq\mathcal{F}}(-1)^{|S_1|+1}q^{|\cap S_1|}\alpha(\beta|\Delta_{\cap S_1})=\sum_{\emptyset\neq S_1\subseteq\mathcal{F}}(-1)^{|S_1|+1}q^{|\cap S_1|}(1-(1-\alpha(\beta|\Delta_{\cap S_1})))=|\Delta_1|-\frac{q}{q-1}\mathcal{A}_{\Delta_1,\beta}$$
and
$$\sum_{\emptyset\neq S_2\subseteq\mathcal{H}}(-1)^{|S_2|+1}q^{|\cap S_2|}\alpha(\beta|\Delta_{\cap S_2})=|\Delta_2|-\frac{q}{q-1}\mathcal{A}_{\Delta_2,\beta}.$$
Therefore, we have
\begin{eqnarray}\label{newT}
\mathcal{T}&=&
|\Delta_1|\mathcal{A}_{\Delta_2,\beta}+\frac{q-1}{q}\Big(|\Delta_1||\Delta_2|-\big(|\Delta_1|-\frac{q}{q-1}\mathcal{A}_{\Delta_1,\beta}\big)\big(|\Delta_2|-\frac{q}{q-1}\mathcal{A}_{\Delta_2,\beta}\big)\Big)\nonumber\\
&=&(2|\Delta_1|-\frac{q}{q-1}\mathcal{A}_{\Delta_1,\beta})\mathcal{A}_{\Delta_2,\beta}+|\Delta_2|\mathcal{A}_{\Delta_1,\beta}.
\end{eqnarray}

(1) If  $\beta\in\cap_{i\in[l]}\Delta_{F_i}^{\bot}$ and  $\beta\in\cap_{j\in[k]}\Delta_{H_j}^{\bot}$, then $\mathcal{A}_{\Delta_1,\beta}=\mathcal{A}_{\Delta_2,\beta}=0$ according to Lemma  \ref{linear-weight}(1). It induces that $\mathcal{T}=0$.

Conversely, assume that $\mathcal{T}=0$. If $\beta\notin\cap_{i\in[l]}\Delta_{F_i}^{\bot}$, then $\mathcal{A}_{\Delta_1,\beta}> 0$ again by Lemma  \ref{linear-weight}(1). From \eqref{newT}, one has that
$\mathcal{T}>|\Delta_2|\mathcal{A}_{\Delta_1,\beta}>0$
since $\mathcal{A}_{\Delta_2,\beta}\geq0$, and
\begin{eqnarray}\label{2D-sumF}
2|\Delta_1|-\frac{q}{q-1}\mathcal{A}_{\Delta_1,\beta}\geq2|\Delta_1|-\sum\nolimits_{i\in[l]}q^{|F_i|}>0
\end{eqnarray}
from Lemma \ref{linear-weight}(3) and Lemma \ref{delta-F1}, which  contradicts to the assumption $\mathcal{T}=0$. If  $\beta\notin\cap_{j\in[k]}\Delta_{H_j}^{\bot},$ one can also obtain that $\mathcal{T}>0$. Therefore, we conclude that $\mathcal{T}=0$ if and only if  $\beta\in\cap_{i\in[l]}\Delta_{F_i}^{\bot}$ and  $\beta\in\cap_{j\in[k]}\Delta_{H_j}^{\bot}$, which is equivalent to $\beta\in \Delta_{(\cup_{i\in[l]}F_i)\cup(\cup_{j\in[k]}H_j)}^{\bot}$ by Lemma \ref{cap-cup-bot}(2).

(2) When $\mathcal{T}\ne0$, we consider the minimum possible value of $\mathcal{T}$. If $\mathcal{A}_{\Delta_2,\beta}=0$, then \eqref{newT} and Lemma \ref{linear-weight}(2) indicate that
$$\mathcal{T}=|\Delta_2|\mathcal{A}_{\Delta_1,\beta}\geq (q-1)|\Delta_2| \mathop{\min}\nolimits_{i\in [l]}q^{|F_i|-1}.$$
If $\mathcal{A}_{\Delta_2,\beta}\ne0$, since $2|\Delta_1|-\frac{q}{q-1}\mathcal{A}_{\Delta_1,\beta}>0$ from  \eqref{2D-sumF}, one then has that when $\mathcal{A}_{\Delta_1,\beta}$ is fixed, the value of $\mathcal{T}$  increases as $\mathcal{A}_{\Delta_2,\beta}$ increases. Thus, in this case, we have
\begin{eqnarray*}\mathcal{T}&\geq&(2|\Delta_1|-\frac{q}{q-1}\mathcal{A}_{\Delta_1,\beta}) (q-1)\mathop{\min}\nolimits_{j\in [k]}q^{|H_j|-1}+|\Delta_2|\mathcal{A}_{\Delta_1,\beta}\\
&=&2(q-1)|\Delta_1|\mathop{\min}\nolimits_{j\in [k]}q^{|H_j|-1}+(|\Delta_2|-\mathop{\min}\nolimits_{j\in [k]}q^{|H_j|})\mathcal{A}_{\Delta_1,\beta}\\
&\geq&2(q-1)|\Delta_1|\mathop{\min}\nolimits_{j\in [k]}q^{|H_j|-1}.
\end{eqnarray*}
 due to $|\Delta_2|\geq\mathop{\min}\nolimits_{j\in [k]}q^{|H_j|}$ and $\mathcal{A}_{\Delta_1,\beta}\geq0$.
 This proves the first part of Lemma \ref{two-linear-weight}(2), and the second part can be easily checked by using Lemma \ref{linear-weight}.

 (3) Due to $\mathcal{A}_{\Delta_1,\beta}\leq(q-1)\sum\nolimits_{i\in[l]}q^{|F_i|-1}$,   $\mathcal{A}_{\Delta_2,\beta}\leq(q-1)\sum\nolimits_{j\in[k]}q^{|H_j|-1}$ and \eqref{2D-sumF}, one has
 \begin{eqnarray*}\mathcal{T}&\leq&(2|\Delta_1|-\frac{q}{q-1}\mathcal{A}_{\Delta_1,\beta}) (q-1)\sum\nolimits_{j\in[k]}q^{|H_j|-1}+|\Delta_2|\mathcal{A}_{\Delta_1,\beta}\\
&=&2(q-1)|\Delta_1|\sum\nolimits_{j\in[k]}q^{|H_j|-1}+(|\Delta_2|-\sum\nolimits_{j\in[k]}q^{|H_j|})\mathcal{A}_{\Delta_1,\beta}\\
&\leq&2(q-1)|\Delta_1|\sum\nolimits_{j\in[k]}q^{|H_j|-1}.
\end{eqnarray*}
The forgoing equality holds if and only if $\mathcal{A}_{\Delta_1,\beta}=0$  and   $\mathcal{A}_{\Delta_2,\beta}=(q-1)\sum\nolimits_{j\in[k]}q^{|H_j|-1}$, which is equivalent to $\beta\notin\cup_{j\in[k]}\Delta_{H_j}^{\bot}$ and $\beta\in\Delta_{( \cup_{1\leq i<j\leq k}H_i\cap H_j)\cup(\cup_{i\in[l]}F_i)}^{\bot}$ from Lemma \ref{linear-weight}(1) and (3). This completes the proof.
\end{IEEEproof}
The following lemma presents the necessary and sufficient conditions on  simplicial complexes under which  some $\beta\in\mathbb{F}_q^m$ satisfies Lemmas \ref{linear-weight} and \ref{two-linear-weight}.

\begin{lem}\label{equi-beta-set} Let $m$, $l$, $k$ be positive integers. Let $F_i,H_j\subseteq[m]$ and $\Delta_{F_i},\Delta_{H_j}$ be  the simplicial complexes of $\mathbb{F}_{q}^m$ for $i\in[l]$ and $j\in[k]$. Then we have
\begin{enumerate}
    \item [(1)] There exists $\beta\in\mathbb{F}_{q}^m$ such that $\beta\notin\Delta_{F_i}^{\bot}$ and  $\beta\in\Delta_{\cup_{j\in[l]\backslash\{i\}}F_j}^{\bot}$ if and only if $F_i\backslash\cup_{j\in[l]\backslash\{i\}}F_j\neq\emptyset$;

   \item [(2)] There exists $\beta\in\mathbb{F}_{q}^m$ such that $\beta\notin\cup_{i\in[l]}\Delta_{F_i}^{\bot}$ and $\beta\in \cap_{1\leq i<j\leq l}\Delta_{F_i\cap F_j}^{\bot}$ if and only if $F_i\backslash\cup_{j\in[l]\backslash\{i\}}F_j\neq\emptyset$ for any $i\in[l]$ .

       \item [(3)] There exists $\beta\in\mathbb{F}_{q}^m$ such that $\beta\notin\Delta_{F_i}^{\bot}$ and  $\beta\in\Delta_{(\cup_{\ell\in[l]\backslash\{i\}}F_\ell)\cup(\cup_{j\in[k]}H_j)}^{\bot}$ if and only if $F_i\backslash\big((\cup_{\ell\in[l]\backslash\{i\}}F_\ell)\cup(\cup_{j\in[k]}H_j)\big)\neq\emptyset$;

   \item [(4)] There exists $\beta\in\mathbb{F}_{q}^m$ such that  $\beta\notin\cup_{i\in[l]}\Delta_{F_i}^{\bot}$  and $\beta\in\Delta_{( \cup_{1\leq i<j\leq l}F_i\cap F_j)\cup(\cup_{j\in[k]}H_j)}^{\bot}$ if and only if $F_i\backslash\big((\cup_{j\in[l]\backslash\{i\}}F_j) \cup(\cup_{j\in[k]}H_j)\big)
   \neq\emptyset$ for any $i\in[l]$.
       \end{enumerate}
\end{lem}

\begin{IEEEproof} (1) When $F_i\backslash\cup_{j\in[l]\backslash\{i\}}F_j\neq\emptyset$, it is clear that if $\text{supp}(\beta)=F_i\backslash\cup_{j\in[l]\backslash\{i\}}F_j$, then $\beta\notin\Delta_{F_i}^{\bot}$ and  $\beta\in\Delta_{\cup_{j\in[l]\backslash\{i\}}F_j}^{\bot}$. Suppose $F_i\backslash\cup_{j\in[l]\backslash\{i\}}F_j=\emptyset$. Then $F_i\cap(\cup_{j\in[l]\backslash\{i\}}F_j)=F_i$ and thus, $\beta\in\Delta_{\cup_{j\in[l]\backslash\{i\}}F_j}^{\bot}$ implies $\beta\in\Delta_{F_i}^{\bot}$ by Lemma \ref{cap-cup-bot}(1). Therefore, there cannot exist $\beta\in\mathbb{F}_{q}^m$ satisfies $\beta\notin\Delta_{F_i}^{\bot}$ and  $\beta\in\Delta_{\cup_{j\in[l]\backslash\{i\}}F_j}^{\bot}$ under this assumption.

(2) If $F_i\backslash\cup_{j\in[l]\backslash\{i\}}F_j\neq\emptyset$ for any $i\in[l]$, then $\beta\in\mathbb{F}_q^m$ with $\text{supp}(\beta)=\cup_{i\in[l]}\left(F_i\backslash\cup_{j\in[l]\backslash\{i\}}F_j\right)$ satisfies $\beta\notin\Delta_{F_i}^{\bot}$ for any $i\in[l]$ and $\beta\in \cap_{1\leq i<j\leq l}\Delta_{F_i\cap F_j}^{\bot}$. Conversely, without loss of generality,  suppose that  $F_1\backslash\cup_{j\in[l]\backslash\{1\}}F_j=\emptyset$. Then   $F_1= F_1\cap(\cup_{j\in[l]\backslash\{1\}}F_j)\subseteq\cup_{1\leq i<j\leq l}(F_i\cap F_j)$. Therefore, one can get that
 $$ \cap_{1\leq i<j\leq l}\Delta_{F_i\cap F_j}^{\bot}=\Delta_{ \cup_{1\leq i<j\leq l}(F_i\cap F_j)}^{\bot}\subseteq\Delta_{F_1}^{\bot}$$
 again by Lemma \ref{cap-cup-bot}, which implies that such $\beta$ cannot exist.

 (3) and (4) can be similarly proved. This completes the proof.
 \end{IEEEproof}
\section{Linear Codes over Non-unital  Non-commutative Rings}
In this section, we shall give the  parameters of $\mathcal{C}_L$ in \eqref{Ring-Trace-Code}  over the ring $\mathcal{R}=E^s$ and $\mathcal{R}=F^s$, respectively, and consider their corresponding subfield-like codes. Specially, we determine their Lee-weight distributions when certain simplicial complexes are employed, which extend the results in \cite{SS}.

\subsection{Codes over Rings and the Corresponding Subfield-like Codes}
 The following conclusions can be derived from Lemmas \ref{value-of-delta}, \ref{linear-weight} and \ref{two-linear-weight} directly.
\begin{thm}\label{thm1}Let $m$, $l$, $s$ be positive integers, $q=p^s$ and $\Delta_1,\Delta_2$ be simplicial complexes of $\mathbb{F}_q^m$ with $\mathcal{F}=\{F_1,\dots,F_l\}$ the set of maximal elements of $\Delta_1$. Let
$ F_i\backslash\cup_{j\in[l]\backslash\{i\}}F_j\neq\emptyset$ for any $i\in[l]$ and $\mathcal{C}_{L}$ be defined in \eqref{Ring-Trace-Code} with $\mathcal{R}=E^s$.
\begin{enumerate}
    \item [(1)]  Let $L=a\Delta_1+c\Delta_2$. Then  $\mathcal{C}_{L}$ is a code over $\mathcal{R}$ of length
    $ |\Delta_1| |\Delta_2|$, size $q^{2|\cup_{i\in[l]}F_i|}$ and  $\min d_L=|\Delta_2| (q-1) \mathop{\min}_{i\in [l]}q^{|F_i|-1}$;

    \item [(2)]  Let $L=a\Delta_1+c\Delta_2^{c}$.
    Then  $\mathcal{C}_{L}$ is a code over $\mathcal{R}$ of length  $|\Delta_1| (q^m-|\Delta_2|)$,  size $q^{2|\cup_{i\in[l]}F_i|}$ and $\min d_L=(q^m-|\Delta_2|) (q-1) \mathop{\min}_{i\in [l]}q^{|F_i|-1}$;

    \item [(3)]  Let $L=a\Delta_1^{c}+c\Delta_2$, $q^m>\sum_{i\in[l]}q^{|F_i|}$. Then $\mathcal{C}_{L}$ is a code over $\mathcal{R}$ of length $(q^m- |\Delta_1|) |\Delta_2|$, size $q^{2m}$ and
$\min d_L=(q-1)|\Delta_2|(q^{m-1}-\sum_{i\in[l]}q^{|F_i|-1})$;

    \item [(4)]  Let $L=a\Delta_1^{c}+c\Delta_2^{c}$, $q^m>\sum_{i\in[l]}q^{|F_i|}$. Then  $\mathcal{C}_{L}$ is a code over $\mathcal{R}$ of length $(q^m- |\Delta_1|)(q^m- |\Delta_2|)$, size $q^{2m}$ and
$\min d_L= (q-1)(q^m- |\Delta_2|)(q^{m-1}-\sum_{i\in[l]}q^{|F_i|-1})$;

    \item [(5)]  Let $L=(a\Delta_1+c\Delta_2)^{c}$ and $q^m>\sum_{i\in[l]}q^{|F_i|}$. Then $\mathcal{C}_{L}$ is a code over $\mathcal{R}$ of length $q^{2m}-|\Delta_1| |\Delta_2|$, size $q^{2m}$ and
 $\min d_L= (q-1)(q^{2m-1}-|\Delta_2|\sum_{i\in[l]}q^{|F_i|-1})$,
  \end{enumerate}
  where $|\Delta_1|$ and $|\Delta_2|$ can be calculated by Lemma \ref{value-of-delta}.
\end{thm}
\begin{IEEEproof}
We only give the proof of (1) and (3) since the others can be proved in a similar manner.

{\em Proof of (1)}:  Putting $L_1=\Delta_1$ and $L_2=\Delta_2$, it is clear that the length of $\mathcal{C}_L$ is $|\Delta_1||\Delta_2|$.
For  any $v=a\beta_1+c\beta_2\in\mathcal{R}^m$, one can derive from  Equation \eqref{weight-for-E} and Lemma \ref{linear-weight} that
\begin{eqnarray}\label{weight-thm1(1)}
wt_L(c_{L}(v))=|\Delta_2|(\mathcal{A}_{\Delta_1,\beta_2}+\mathcal{A}_{\Delta_1,\beta_1+\beta_2})
\end{eqnarray}
Recall that $\mathcal{A}_{\Delta_1,\beta}$ is the Hamming weight of $c_{\Delta_1}(\beta)$ in $\mathcal{C}_{\Delta_1}$ defined by \eqref{D-Ding}. Therefore, we have
 $wt_L(c_{L}(v))=0$ if and only if  $\mathcal{A}_{\Delta_1,\beta_2}=\mathcal{A}_{\Delta_1,\beta_1+\beta_2}=0.$
 Further, by Lemma \ref{linear-weight}(1), one has $wt_L(c_{L}(v))=0$ if and only if   $\beta_2\in\Delta_{\cup_{i\in[l]}F_i}^{\bot}$ and $\beta_1+\beta_2\in\Delta_{\cup_{i\in[l]}F_i}^{\bot}$.
  Due to  the independence of $\beta_1$ and $\beta_2$,  the number of $(\beta_1,\beta_2)\in(\mathbb{F}_{q}^m)^2$, which satisfies $wt_L(c_{L}(v))=0$,  is $q^{2(m-|\cup_{i=1}^lF_i|)}$. It induces that the size of $\mathcal{C}_L$ is $q^{2|\cup_{i=1}^lF_i|}$.
  In what follows, we determine the non-zero minimum  Lee weight of $\mathcal{C}_L$, which can exist only if $\beta_2\notin \Delta_{F_i}^{\bot}$ or  $\beta_1+\beta_2\notin \Delta_{F_i}^{\bot}$.

  If  $\beta_2\notin \Delta_{F_i}^{\bot}$ (or  $\beta_1+\beta_2\notin \Delta_{F_i}^{\bot}$) for some $i\in[l]$, then $\mathcal{A}_{\Delta_1,\beta_2}\geq(q-1)q^{|F_i|-1}$ (or $\mathcal{A}_{\Delta_1,\beta_1+\beta_2}\geq(q-1)q^{|F_i|-1}$) by  Lemma  \ref{linear-weight}(2). Moreover,   the both equalities can be achieved since   Lemma \ref{equi-beta-set}(1) and $F_i\backslash(\cup_{j\in[l]\backslash\{i\}}F_j)\neq\emptyset$ for any $i\in[l]$.  It implies that there exists  $(\beta_1,\beta_2)\in(\mathbb{F}_q^m)^2$ such that  $\mathcal{A}_{\Delta_1,\beta_2}=(q-1)q^{|F_i|-1}$, $\mathcal{A}_{\Delta_1,\beta_1+\beta_2}=0$ (or $\mathcal{A}_{\Delta_1,\beta_2}=0$, $\mathcal{A}_{\Delta_1,\beta_1+\beta_2}=(q-1)q^{|F_i|-1}$) due to the independence of $\beta_1, \beta_2$.  Therefore,  the  non-zero minimum Lee weight  of $\mathcal{C}_L$ is   $|\Delta_2| (q-1) \mathop{\min}_{i\in [l]}q^{|F_i|-1}$.

{\em Proof of (3)}:  Let $v=a\beta_1+c\beta_2\in\mathcal{R}^m$ and $(\beta_1,\beta_2)\in(\mathbb{F}_q^m)^2$. If $L=a\Delta_1^{c}+c\Delta_2$, then $|L|=|\Delta_1^{c}||\Delta_2|=(q^m- |\Delta_1|) |\Delta_2|$. Moreover, \eqref{weight-for-E}  and Lemmas \ref{a-simplicial-comp-exponent-sum}, \ref{linear-weight} give that
\begin{eqnarray}\label{weight-thm1-3} wt_L(c_{L}(v))&=&|\Delta_2|\Big(2|\Delta_1^c|-\frac{1}{q}\Big(\sum_{u\in\mathbb{F}_p^s}\sum_{t_1\in \Delta_1^c}w_p^{\langle u,\langle\beta_2,t_1\rangle_q\rangle_p}+\sum_{u\in\mathbb{F}_p^s}\sum_{t_1\in \Delta_1^c}w_p^{\langle u,\langle\beta_1+\beta_2,t_1\rangle_q\rangle_p}\Big)\Big)\nonumber\\
&=&|\Delta_2|\Big(2|\Delta_1^c|-\frac{1}{q}(2q^m+q^m(q-1)(\delta_{0,\beta_2}+\delta_{0,\beta_1+\beta_2}))\Big)\nonumber\\
&&+2|\Delta_1||\Delta_2|-|\Delta_2|(\mathcal{A}_{\Delta_1,\beta_2}+\mathcal{A}_{\Delta_1,\beta_1+\beta_2})\nonumber\\
&=&(q-1)|\Delta_2|q^{m-1}(2-\delta_{0,\beta_2}-\delta_{0,\beta_1+\beta_2})-|\Delta_2|(\mathcal{A}_{\Delta_1,\beta_2}+\mathcal{A}_{\Delta_1,\beta_1+\beta_2})
\end{eqnarray}
due to  $|\Delta_1^c|=q^m-|\Delta_1|$. When $(\beta_{1}+\beta_2,\beta_2)=(0,0)$, one has $wt_L(c_{L}(v))=0$. If $\beta_2(\beta_1+\beta_2)=0$ and $(\beta_{1}+\beta_2,\beta_2)\ne(0,0)$, then there is at least one of $\mathcal{A}_{\Delta_1,\beta_2}$ and $\mathcal{A}_{\Delta_1,\beta_1+\beta_2}$ equals 0 by Lemma \ref{linear-weight}(1). Therefore, $$wt_L(c_{L}(v))=(q-1)|\Delta_2|q^{m-1}-|\Delta_2|(\mathcal{A}_{\Delta_1,\beta_2}+\mathcal{A}_{\Delta_1,\beta_1+\beta_2})\geq (q-1)|\Delta_2|\Big(q^{m-1}-\sum\nolimits_{i\in[l]}q^{|F_i|-1}\Big)>0$$
due to   $\mathcal{A}_{\Delta_1,\beta_2},\mathcal{A}_{\Delta_1,\beta_1+\beta_2}\leq\sum_{i\in[l]}(q-1)q^{|F_i|-1}$ from Lemma \ref{linear-weight}(3) and $q^m>\sum_{i\in[l]}q^{|F_i|}$. If $\beta_2(\beta_1+\beta_2)\ne0$, one can also obtain that
$$wt_L(c_L(v))=2(q-1)|\Delta_2|q^{m-1}-|\Delta_2|(\mathcal{A}_{\Delta_1,\beta_2}+\mathcal{A}_{\Delta_1,\beta_1+\beta_2})\geq 2(q-1)|\Delta_2|\Big(q^{m-1}-\sum\nolimits_{i\in[l]}q^{|F_i|-1}\Big)>0$$
again by Lemma \ref{linear-weight}(3).
Therefore, $wt_L(c_L(v))=0$  only if $(\beta_1,\beta_2)=(0,0)$, which means  that the size of $\mathcal{C}_{L}$ is $q^{2m}$. Moreover, since  $F_i\backslash\cup_{j\in[l]\backslash\{i\}}F_j\neq\emptyset$ for any $i\in[l]$, then  both the forgoing equalities can be achieved  due to  Lemmas \ref{linear-weight}(3) and \ref{equi-beta-set}(2). This induces that non-zero minimum Lee weight of $\mathcal{C}_{L}$ is $(q-1)|\Delta_2|(q^{m-1}-\sum_{i\in[l]}q^{|F_i|-1})$.
\end{IEEEproof}

\begin{thm}\label{Thm2}Let $m$, $l$, $k$ be positive integers, and  $\Delta_1$,  $\Delta_2$ be two simplicial complexes of $\mathbb{F}_q^m$ with $\mathcal{F}=\{F_1,\dots,F_l\}$,  $\mathcal{H}=\{H_1,\dots,H_k\}$ the sets of maximal elements of $\Delta_1$ and  $\Delta_2$, respectively.  Let
 $F_i\backslash((\cup_{\ell\in[l]\backslash\{i\}}F_{\ell})\cup(\cup_{t\in[k]}H_{t}))\neq\emptyset$ and  $H_j\backslash((\cup_{t\in[k]\backslash\{j\}}H_t)\cup(\cup_{\ell\in[l]}F_\ell))\neq\emptyset$ for any $i\in[l]$, $j\in[k]$, and $\mathcal{C}_{L}$ be defined in \eqref{Ring-Trace-Code} with $\mathcal{R}=F^s$.
\begin{enumerate}
    \item [(1)]  Let $L=a\Delta_1+c\Delta_2$. Then $\mathcal{C}_{L}$ is a code over $\mathcal{R}$ of length $ |\Delta_1| |\Delta_2|$, size $q^{ |(\cup_{i\in[l]}F_i)\cup(\cup_{j\in[k]}H_j)|}$ and
$\min d_L=(q-1) \mathop{\min}_{i\in [l],j\in[k]}\{|\Delta_2|q^{|F_i|-1},2|\Delta_1|q^{|H_j|-1}\}$;

    \item [(2)]  Let $L=a\Delta_1+c\Delta_2^{c}$ and   $q^{m}>\sum_{j\in[k]}q^{|H_j|}$. Then $\mathcal{C}_{L}$ is a code over $\mathcal{R}$ of length $ |\Delta_1|(q^m- |\Delta_2|)$, size $q^m$ and
$ \min d_L=2(q-1)|\Delta_1|(q^{m-1}-\sum_{j\in[k]}q^{|H_j|-1})$;

    \item [(3)]  Let $L=a\Delta_1^{c}+c\Delta_2$  and   $q^{m}|\Delta_2|>\kappa_1$, where $\kappa_1=\max\{|\Delta_2|\sum_{i\in[l]}q^{|F_i|},(2|\Delta_1|-q^m)\sum_{j\in[k]}q^{|H_j|}\}$.  Then  $\mathcal{C}_{L}$ is a  code over $\mathcal{R}$ of length
$(q^m- |\Delta_1|) |\Delta_2|$,  size $q^m$ and
$\min d_L=\frac{q-1}{q}(q^{m}|\Delta_2|-\kappa_1)$;

    \item [(4)]  Let $L=a\Delta_1^{c}+c\Delta_2^{c}$, $|\Delta_1|\leq|\Delta_1^c|$ and $q^{m}(|\Delta_1^c|-|\Delta_1|+|\Delta_2^c|)>\kappa_2$, where $\kappa_2=(|\Delta_1^c|-|\Delta_1|)\sum_{j\in[k]}q^{|H_j|}+(\sum_{j\in[k]}q^{|H_j|}-|\Delta_2|)\sum_{i\in[l]}q^{|F_i|}$. Then   $\mathcal{C}_{L}$ is a  code over $\mathcal{R}$ of length
$(q^m- |\Delta_1|)(q^m- |\Delta_2|)$,  size $q^m$ and
$\min d_L=\frac{q-1}{q}(q^{m}(|\Delta_1^c|-|\Delta_1|+|\Delta_2^c|)-\kappa_2)$;

\item [(5)]  Let $L=(a\Delta_1+c\Delta_2)^{c}$, $q^{2m}-|\Delta_1|\sum_{j\in[k]}q^{|H_j|}>0$.  Then   $\mathcal{C}_{L}$ is a  code over $\mathcal{R}$ of length
 $ q^{2m}-|\Delta_1| |\Delta_2|$, size $q^m$ and $\min d_L=2(q-1)(q^{2m-1}-|\Delta_1|\sum_{j\in[k]}q^{|H_j|-1})$,

  \end{enumerate}
   where $|\Delta_1|$ and $|\Delta_2|$ can be calculated by Lemma \ref{value-of-delta}.
\end{thm}

\begin{IEEEproof}We only give the proofs of (1),  (2) and (4) since the others can be proved in a similar manner.

{\em Proof of (1)}:  Letting $L_1=\Delta_1$ and $L_2=\Delta_2$, it is clear that the length of $\mathcal{C}_L$ is $|\Delta_1||\Delta_2|$.
For  any $v=a\beta_1+c\beta_2\in\mathcal{R}^m$, one can derive from  Equation \eqref{weight-for-F}, Lemmas  \ref{linear-weight} and  \ref{two-linear-weight} that
\begin{eqnarray*}
wt_L(c_{L}(v))=|\Delta_1|\mathcal{A}_{\Delta_2,\beta_1}+\tilde{\mathcal{A}}_{\beta_1}.
\end{eqnarray*}
Note that $wt_L(c_{L}(v))$ is actually the value of $\mathcal{T}$ defined in Lemma \ref{two-linear-weight} for $\beta=\beta_1$.  Therefore,  $wt_L(c_{L}(v))=0$ if and only if $\beta_1\in \Delta_{(\cup_{i\in[l]}F_i)\cup(\cup_{j\in[k]}H_j)}^{\bot}$. It induces that the size of $\mathcal{C}_L$ is $q^{|(\cup_{i\in[l]}F_i)\cup(\cup_{j\in[k]}H_j)|}$.

If $wt_L(c_{L}(v))\ne0$, then Lemma \ref{two-linear-weight}(2) gives that
$$wt_L(c_{L}(v))\geq (q-1) \mathop{\min}\nolimits_{i\in [l],j\in[k]}\{|\Delta_2|q^{|F_i|-1},2|\Delta_1|q^{|H_j|-1}\}.$$
Since  $F_i\backslash((\cup_{\ell\in[l]\backslash\{i\}}F_\ell)\cup(\cup_{t\in[k]}H_t))\neq\emptyset$ and $H_j\backslash((\cup_{t\in[k]\backslash\{j\}}H_t)\cup(\cup_{\ell\in[l]}F_\ell))\neq\emptyset$ for any $i\in[l]$ and $j\in[k]$,  by  Lemma  \ref{two-linear-weight}(2) and Lemma \ref{equi-beta-set}(3), it can be seen that the all the values in the set $(q-1)\{|\Delta_2|q^{|F_i|-1},2|\Delta_1|q^{|H_j|-1}: i\in [l],j\in[k]\}$ can be achieved. Therefore, we can conclude that
$$\min d_L=(q-1) \mathop{\min}\nolimits_{i\in [l],j\in[k]}\{|\Delta_2|q^{|F_i|-1},2|\Delta_1|q^{|H_j|-1}\}.$$

{\em Proof of (2)}: If  $L=a\Delta_1+c\Delta_2^{c}$, then for  any $v=a\beta_1+c\beta_2\in\mathcal{R}^m$, \eqref{weight-for-F} and Lemmas \ref{a-simplicial-comp-exponent-sum}-\ref{two-linear-weight} give that
\begin{eqnarray}\label{thm2(2)-weight}
wt_L(c_{L}(v))=2(q-1)q^{m-1}|\Delta_1|(1-\delta_{0,\beta_1})-(|\Delta_1|\mathcal{A}_{\Delta_2,\beta_1}+\tilde{\mathcal{A}}_{\beta_1}).
\end{eqnarray}
Since
\begin{eqnarray*}
|\Delta_1|\mathcal{A}_{\Delta_2,\beta_1}+\tilde{\mathcal{A}}_{\beta_1}\leq 2(q-1)|\Delta_1|\sum\nolimits_{j\in[k]}q^{|H_j|-1}
\end{eqnarray*}
from Lemma \ref{two-linear-weight}(3).
 one then get  that for any $\beta_1\ne 0$,
$$wt_L(c_{L}(v))=2(q-1)q^{m-1}|\Delta_1|-(|\Delta_1|\mathcal{A}_{\Delta_2,\beta_1}+\tilde{\mathcal{A}}_{\beta_1})\geq 2(q-1)|\Delta_1|(q^{m-1}-\sum\nolimits_{j\in[k]}q^{|H_j|-1})>0.$$
Moreover, $wt_L(c_{L}(v))= 2(q-1)|\Delta_1|(q^{m-1}-\sum\nolimits_{j\in[k]}q^{|H_j|-1})$ if and only if $$\beta_1\notin\cup_{j\in[k]}\Delta_{H_j}^{\bot}\;\text{and}\;\beta_1\in\Delta_{( \cup_{1\leq i<j\leq k}H_i\cap H_j)\cup(\cup_{i\in[l]}F_i)}^{\bot}.$$
According to Lemma \ref{equi-beta-set}(4) and the assumption  $H_j\backslash((\cup_{t\in[k]\backslash\{j\}}H_t)\cup(\cup_{\ell\in[l]}F_\ell))\neq\emptyset$ for any $j\in[k]$, we can see that such $\beta_1$ exists. This completes the proof.

{\em Proof of (4)}: If  $L=a\Delta_1^c+c\Delta_2^{c}$, then  \eqref{weight-for-F} and Lemmas \ref{a-simplicial-comp-exponent-sum}-\ref{two-linear-weight} give that for  any $v=a\beta_1+c\beta_2\in\mathcal{R}^m$ with $\beta_1\ne0$,
\begin{eqnarray}\label{thm2(4)-weight}
wt_L(c_L(v))&=&(q-1)q^{m-1}(|\Delta_1^c|+|\Delta_2^c|-|\Delta_1|)-|\Delta_1^c|\mathcal{A}_{\Delta_2,\beta_1}
+\tilde{\mathcal{A}}_{\beta_1}\nonumber\\
&=&(q-1)q^{m-1}(|\Delta_1^c|+|\Delta_2^c|-|\Delta_1|)+V
\end{eqnarray}
from \eqref{newT},
where $$V=\Big(|\Delta_1|-|\Delta_1^c|-\frac{q}{q-1}\mathcal{A}_{\Delta_1,\beta_1}\Big)\mathcal{A}_{\Delta_2,\beta_1}
+|\Delta_2|\mathcal{A}_{\Delta_1,\beta_1}.$$ Since $|\Delta_1^c|\geq|\Delta_1|$ which implies $|\Delta_1^c|+|\Delta_2^c|-|\Delta_1|\geq0$,  we need to consider the minimum value of $V$ when $\beta_1$ runs through  $\mathbb{F}_q^m\backslash\{0\}$.  Observe that
when $\mathcal{A}_{\Delta_1,\beta_1}$ is fixed, the value of $V$ decreases as $\mathcal{A}_{\Delta_2,\beta_1}$ increases due to $|\Delta_1|-|\Delta_1^c|-\frac{q}{q-1}\mathcal{A}_{\Delta_1,\beta_1}\leq0$. From the fact that  $\mathcal{A}_{\Delta_2,\beta_1}\leq(q-1)\sum_{j\in[k]}q^{|H_j|-1}$ by Lemma \ref{linear-weight}(3), we have
\begin{eqnarray*}
 V&\geq & \Big(|\Delta_1|-|\Delta_1^c|-\frac{q}{q-1}\mathcal{A}_{\Delta_1,\beta_1}\Big)(q-1)\sum\nolimits_{j\in[k]}q^{|H_j|-1}
+|\Delta_2|\mathcal{A}_{\Delta_1,\beta_1}\\
&=&\Big(|\Delta_2|-\sum\nolimits_{j\in[k]}q^{|H_j|}\Big)\mathcal{A}_{\Delta_1,\beta_1}
+(|\Delta_1|-|\Delta_1^c|)(q-1)\sum\nolimits_{j\in[k]}q^{|H_j|-1}\\
&\geq&\frac{q-1}{q}\Big(\big(|\Delta_2|-\sum\nolimits_{j\in[k]}q^{|H_j|}\big)\sum\nolimits_{i\in[l]}q^{|F_i|}
+\left(|\Delta_1|-|\Delta_1^c|\right)\sum\nolimits_{j\in[k]}q^{|H_j|}\Big)
\end{eqnarray*}
due to $|\Delta_2|-\sum\nolimits_{j\in[k]}q^{|H_j|}\leq 0$ and $\mathcal{A}_{\Delta_1,\beta_1}\leq (q-1)\sum\nolimits_{i\in[l]}q^{|F_i|-1}$ again by Lemma \ref{linear-weight}(3). Further, one has
$$wt_L(c_L(v))\geq\frac{q-1}{q}\Big(q^m(|\Delta_1^c|+|\Delta_2^c|-|\Delta_1|)-
\Big(\sum\nolimits_{j\in[k]}q^{|H_j|}-|\Delta_2|\Big)\sum\nolimits_{i\in[l]}q^{|F_i|}
-\left(|\Delta_1^c|-|\Delta_1|\right)\sum\nolimits_{j\in[k]}q^{|H_j|}\Big).$$
 Moreover, the forgoing equality holds if and only if
$$\mathcal{A}_{\Delta_1,\beta_1}=(q-1)\sum\nolimits_{i\in[l]}q^{|F_i|-1}\quad {\text {and}}\quad
 \mathcal{A}_{\Delta_2,\beta_1}=(q-1)\sum\nolimits_{j\in[k]}q^{|H_j|-1}.$$
 When we select
$\beta_1\in\mathbb{F}_q^m\backslash\{0\}$ with the form
$$\text{supp}(\beta_1)=\cup_{i\in[l]}F_i\backslash\left((\cup_{\ell\in[l]\backslash\{i\}}F_\ell)\cup(\cup_{t\in[k]}H_t)\right)\cup
\left(\cup_{j\in[k]}H_j\backslash\left((\cup_{t\in[k]\backslash\{j\}}H_t)\cup(\cup_{\ell\in[l]}F_\ell)\right)\right),$$
one can conclude from Lemmas \ref{linear-weight}(3)  and \ref{equi-beta-set}(2)  that the equality can be achieved due to  $F_i\backslash((\cup_{\ell\in[l]\backslash\{i\}}F_\ell)\cup(\cup_{t\in[k]}H_t))\neq\emptyset$ and  $H_j\backslash((\cup_{t\in[k]\backslash\{j\}}H_t)\cup(\cup_{\ell\in[l]}F_\ell))\neq\emptyset$ for any $i\in[l]$, $j\in[k]$. This completes the proof.
\end{IEEEproof}

Next, we aim to construct subfield-like codes \cite{SS1} from $\mathcal{C}_L$ in Theorems \ref{thm1} and \ref{Thm2}. Define the $\mathbb{F}_q$-linear functional $\varphi: \mathcal{R}\rightarrow \mathbb{F}_q$ as $\varphi(ax+cy)=x$, where $x,y\in \mathbb{F}_q$. For $m\in\mathbb{N}$, the mapping $\varphi$ can be extended naturally from $\mathcal{R}^m$ to $\mathbb{F}_{q}^{m}$ as $\varphi(ax+cy)=x$, where $x,y\in \mathbb{F}_q^m$. By  \eqref{Ring-Trace-Code}, we have that
$$\varphi(\mathcal{C}_{L})=\{\varphi(c_{L}(v))=(\varphi(\langle v, x\rangle_q))_{x\in L}: v \in \mathcal{R}^m\}\subseteq\mathbb{F}_q^{|L|},$$
and $\varphi(\mathcal{C}_{L})$ is a linear subfield-like code over $\mathbb{F}_q$.  Assume that $v=a\beta_1+c\beta_2\in\mathcal{R}^m$ and $x=at_1+ct_2\in L$, where $\beta_i\in\mathbb{F}_{q}^m$ and $t_i\in L_i$, $i=1,2$. Then by \eqref{vector-product}, one has
$$wt_H(\varphi(c_{L}(v)))=wt_H(\varphi(\langle v, x\rangle_q))_{x\in L})=wt_H((\langle \beta_1, t_1\rangle_q)_{t_1\in L_1,t_2\in L_2})$$
no matter $\mathcal{R}=E^s$ or $\mathcal{R}=F^s$. Observe that $wt_H(\varphi(c_{L}(v)))$ has the same form of the first part of $wt_L(c_{L}(v))$  in \eqref{weight-for-E}. Therefore, the value of $wt_H(\varphi(c_{L}(v)))$ can be calculated with the help of Lemmas \ref{value-of-delta}, \ref{a-simplicial-comp-exponent-sum} and \ref{linear-weight} as the calculation of the Lee weight of $\mathcal{C}_{L}$ in Theorem \ref{thm1}. Then we have the following result and we omit the proof here.

\begin{thm}\label{subfield-like}Let $m$, $l$, $s$ be positive integers, $q=p^s$ and $\Delta_1,\Delta_2$ be simplicial complexes of $\mathbb{F}_q^m$ with $\mathcal{F}=\{F_1,\dots,F_l\}$ the set of maximal elements of $\Delta_1$. Let
$ F_i\backslash\cup_{j\in[l]\backslash\{i\}}F_j\neq\emptyset$ for any $i\in[l]$ and $\mathcal{C}_{L}$ be defined in \eqref{Ring-Trace-Code} with $\mathcal{R}=E^s$ or $\mathcal{R}=F^s$.
\begin{enumerate}
    \item [(1)]  Let $L=a\Delta_1+c\Delta_2$. Then  $\varphi(\mathcal{C}_{L})$ is a
    $[ |\Delta_1| |\Delta_2|, |\cup_{i\in[l]}F_i|,|\Delta_2| (q-1) \mathop{\min}_{i\in [l]}q^{|F_i|-1}]$ linear code over $\mathbb{F}_q$;

    \item [(2)]  Let $L=a\Delta_1+c\Delta_2^{c}$.
    Then $\varphi(\mathcal{C}_{L})$ is a  $[|\Delta_1| (q^m-|\Delta_2|),  |\cup_{i\in[l]}F_i|,(q^m-|\Delta_2|) (q-1) \mathop{\min}_{i\in [l]}q^{|F_i|-1}]$  linear code over $\mathbb{F}_q$;

    \item [(3)]  Let $L=a\Delta_1^{c}+c\Delta_2$ and $q^m>\sum_{i\in[l]}q^{|F_i|}$. Then $\varphi(\mathcal{C}_{L})$  is a $[(q^m- |\Delta_1|) |\Delta_2|,m,(q-1)|\Delta_2|(q^{m-1}-\sum_{i\in[l]}q^{|F_i|-1})]$  linear code over $\mathbb{F}_q$;

    \item [(4)]  Let $L=a\Delta_1^{c}+c\Delta_2^{c}$ and $q^m>\sum_{i\in[l]}q^{|F_i|}$. Then $\varphi(\mathcal{C}_{L})$  is a  $[(q^m- |\Delta_1|)(q^m- |\Delta_2|), m, (q-1)(q^m- |\Delta_2|)(q^{m-1}-\sum_{i\in[l]}q^{|F_i|-1})]$   linear code over $\mathbb{F}_q$;

    \item [(5)]  Let $L=(a\Delta_1+c\Delta_2)^{c}$ and $q^m>\sum_{i\in[l]}q^{|F_i|}$. Then $\varphi(\mathcal{C}_{L})$  is a $[q^{2m}-|\Delta_1| |\Delta_2|,m, (q-1)(q^{2m-1}-|\Delta_2|\sum_{i\in[l]}q^{|F_i|-1})]$   code over $\mathbb{F}_q$,
  \end{enumerate}
  where $|\Delta_1|$ and $|\Delta_2|$ can be calculated by Lemma \ref{value-of-delta}.
\end{thm}

\begin{remark} In Theorems \ref{thm1}-\ref{subfield-like}, we consider the simplicial complexes $\Delta_1$ and $\Delta_2$ which have arbitrary number of maximal elements. By employing the dual of subspace and the orthogonality of exponential sums, we give  a characterization on the values of $\mathcal{A}_{\Delta,\beta}$ and $\mathcal{T}$, which are closely related to the  weight of codewords in Theorems \ref{thm1}-\ref{subfield-like} (see Lemmas \ref{linear-weight} and \ref{two-linear-weight} for details). One can see that the method used in this paper can be applied into many other rings. Therefore, these results give the positive answer to the first question in Introduction B.
\end{remark}

\subsection{Explicit Lee Weight Distribution of $\mathcal{C}_L$ When $l=k=1$}
In this subsection, we shall present the Lee weight distribution of $\mathcal{C}_L$ when $l=k=1$ in Theorems \ref{thm1} and \ref{Thm2}. When $l,k>1$, one can also obtain the explicit Lee weight distribution of $\mathcal{C}_L$ with the similar method and a routine computation step by step.

\begin{prop}\label{thm3}Let $s$, $m$ be  positive integers, $q=p^s$,  $\mathcal{R}=E^s$ and $ A,B\subseteq[m]$ with $A\ne \emptyset$. Let $\mathcal{C}_{L}$ be defined in \eqref{Ring-Trace-Code}.
\begin{enumerate}
    \item [(1)]  Let $L=a\Delta_A+c\Delta_B$. Then $\mathcal{C}_{L}$  has length $q^{|A|+|B|}$, size $ q^{2|A|}$ and the  Lee weight enumerator
 $$ 1+2(q^{|A|}-1)x^{(q-1)q^{|A|+|B|-1}}+(q^{|A|}-1)^2x^{2(q-1)q^{|A|+|B|-1}}.$$

    \item [(2)]  Let $L=a\Delta_A+c\Delta_B^{c}$. Then  $\mathcal{C}_{L}$ has length
$q^{|A|}(q^m-q^{|B|})$,  size $q^{2|A|}$ and the  Lee weight enumerator
 $$ 1+2(q^{|A|}-1)x^{(q-1)q^{|A|-1}(q^m-q^{|B|})}+(q^{|A|}-1)^2x^{2(q-1)q^{|A|-1}(q^m-q^{|B|})}.$$

    \item [(3)]  Let $L=a\Delta_A^{c}+c\Delta_B$ and $|A|<m$. Then  $\mathcal{C}_{L}$ has length
$q^{|B|}(q^m-q^{|A|})$, size $ q^{2m}$ and the Lee weight distribution in TABLE \ref{Lee weightin Theorem1(3)}.

    \item [(4)]  Let $L=a\Delta_A^{c}+c\Delta_B^{c}$ and $|A|<m$. Then  $\mathcal{C}_{L}$ has length
$(q^{m}-q^{|A|})(q^m-q^{|B|})$, size $ q^{2m}$ and the  Lee weight distribution in TABLE \ref{Lee weightin Theorem1(4)}.

    \item [(5)]  Let $L=(a\Delta_A+c\Delta_B)^{c}$ and $|A|<m$. $\mathcal{C}_{L}$ has length
 $q^{2m}-q^{|A|+|B|}$, size $ q^{2m}$ and the Lee weight distribution in TABLE \ref{Lee weightin Theorem1(5)}.
  \end{enumerate}
\end{prop}

\begin{table}[!htb]
\footnotesize
\centering
\renewcommand{\arraystretch}{0.8}
\caption{Lee weight distribution in Proposition \ref{thm1}(3)}\label{Lee weightin Theorem1(3)}
\setlength{\tabcolsep}{6mm}{
\begin{tabular}{|c|c|}
\hline
 Lee weight &  Frequency\\
\hline\hline
0 & 1\\ \hline
$(q-1)q^{m+|B|-1}$&$2(q^{m-|A|}-1)$\\ \hline
$(q-1)q^{|B|-1}(q^m-q^{|A|})$ &$2(q^m-q^{m-|A|})$\\ \hline
$(q-1)q^{|B|-1}(2q^m-q^{|A|}) $&$2(q^{m-|A|}-1)(q^m-q^{m-|A|})$\\ \hline
$2(q-1)q^{|B|-1}(q^m-q^{|A|}) $&$(q^m-q^{m-|A|})^2$\\ \hline
$2(q-1)q^{m+|B|-1}$ &$(q^{m-|A|}-1)^2$\\ \hline
\end{tabular}}
\end{table}

\begin{table}[!htb]
\footnotesize
\centering
\renewcommand{\arraystretch}{0.8}
\caption{Lee weight distribution in Proposition \ref{thm1}(4)}\label{Lee weightin Theorem1(4)}
\setlength{\tabcolsep}{6mm}{
\begin{tabular}{|c|c|}
\hline
 Lee weight &  Frequency\\
\hline\hline
0 & 1\\ \hline
$(q-1)q^{m-1}(q^{m}-q^{|B|})$&$2(q^{m-|A|}-1)$\\ \hline
$(q-1)(q^{m}-q^{|A|})(q^{m-1}-q^{|B|-1})$ &$2(q^m-q^{m-|A|})$\\ \hline
$(q-1)(2q^{m}-q^{|A|})(q^{m-1}-q^{|B|-1}) $&$2(q^{m-|A|}-1)(q^m-q^{m-|A|})$\\ \hline
$2(q-1)(q^{m}-q^{|A|})(q^{m-1}-q^{|B|-1}) $&$(q^m-q^{m-|A|})^2$\\ \hline
$2(q-1)q^{m-1}(q^{m}-q^{|B|})$ &$(q^{m-|A|}-1)^2$\\ \hline
\end{tabular}}
\end{table}

\begin{table}[!htb]
\footnotesize
\centering
\renewcommand{\arraystretch}{0.8}
\caption{Lee weight distribution in Proposition \ref{thm1}(5)}\label{Lee weightin Theorem1(5)}
\setlength{\tabcolsep}{6mm}{
\begin{tabular}{|c|c|}
\hline
 Lee weight &  Frequency\\
\hline\hline
0 & 1\\ \hline
$(q-1)q^{2m-1}$&$2(q^{m-|A|}-1)$\\ \hline
$(q-1)(q^{2m-1}-q^{|A|+|B|-1})$ &$2(q^m-q^{m-|A|})$\\ \hline
$(q-1)(2q^{2m-1}-q^{|A|+|B|-1})$&$2(q^{m-|A|}-1)(q^m-q^{m-|A|})$\\ \hline
$2(q-1)(q^{2m-1}-q^{|A|+|B|-1})$&$(q^m-q^{m-|A|})^2$\\ \hline
$2(q-1)q^{2m-1}$ &$(q^{m-|A|}-1)^2$\\ \hline
\end{tabular}}
\end{table}

\begin{IEEEproof}We only discuss the proof of part (3). The other parts can be proved in a similar way. It is clear that $|L|=|\Delta_A^{c}||\Delta_B|=q^{|B|}(q^m-q^{|A|})$. Let $v=a\beta_1+c\beta_2\in\mathcal{R}^m$ with $(\beta_1,\beta_2)\in(\mathbb{F}_q^m)^2$, \eqref{weight-thm1-3} and  Lemma \ref{linear-weight} show that
\begin{eqnarray}\label{weight--thm1-3-1}wt_L(c_L(v))=(q-1)q^{|B|-1}\left(q^m(2-\delta_{0,\beta_2}-\delta_{0,\beta_1+\beta_2})-q^{|A|}(2-\alpha(\beta_2|\Delta_A)-\alpha(\beta_1+\beta_2|\Delta_A)\right).
\end{eqnarray}
It can be easily obtained that $wt_L(c_L(v))=0$ when $(\beta_2,\beta_1+\beta_2)=(0,0)$.

When  $\beta_2=0$ and $\beta_1+\beta_2\ne 0$, then $wt_L(c_L(v))=(q-1)q^{|B|-1}(q^m-q^{|A|}(1-\alpha(\beta_1|\Delta_A))$ by \eqref{weight--thm1-3-1}. Further,  one has $wt_L(c_L(v))=(q-1)q^{m+|B|-1}$ if $\beta_1\in\Delta_A^{\bot}$, and in this case, the number of $v$ with $wt_L(c_L(v))=(q-1)q^{m+|B|-1}$ is $q^{m-|A|}-1$ due to  $|\Delta_A^{\bot}|=q^{m-|A|}$. Otherwise, if $\beta_1\notin\Delta_A^{\bot}$, then $wt_L(c_L(v))=(q-1)q^{|B|-1}(q^m-q^{|A|})$. Moreover, one can derive that the number of such $v$ is $q^m-q^{m-|A|}$ due to  $|\mathbb{F}_q^m\backslash\Delta_A^{\bot}|=q^m-q^{m-|A|}$ in this case.

When  $\beta_2\ne0$ and $\beta_1+\beta_2= 0$,  one can similarly obtain that $wt_L(c_L(v))=(q-1)q^{m+|B|-1}$ if $\beta_2\in\Delta_A^{\bot}$. In this case, we have  the number of  $v$ is $q^{m-|A|}-1$ since $|\Delta_A^{\bot}|=q^{m-|A|}$ and $\beta_1$ is uniquely determined by $\beta_2$. Otherwise, if $\beta_2\notin\Delta_A^{\bot}$, then $wt_L(c_L(v))=(q-1)q^{|B|-1}(q^m-q^{|A|})$. Moreover, one can derive that the number of  $v$ is $q^m-q^{m-|A|}$ again by $|\mathbb{F}_q^m\backslash\Delta_A^{\bot}|=q^m-q^{m-|A|}$.

When  $\beta_2\ne0 $ and $\beta_1+\beta_2\ne 0$, $wt_L(c_L(v))=(q-1)q^{|B|-1}(2q^m-q^{|A|}(2-\alpha(\beta_2|\Delta_A)-\alpha(\beta_1+\beta_2|\Delta_A))$ by \eqref{weight--thm1-3-1}. If $\beta_2,\beta_1+\beta_2\in\Delta_A^{\bot}$,
then $wt_L(c_L(v))=2(q-1)q^{m+|B|-1}$. In this case, we have the number of  $v$ is $(q^{m-|A|}-1)^2$. If $\beta_2\in\Delta_A^{\bot}, \beta_1+\beta_2\notin\Delta_A^{\bot}$ or $\beta_2\notin\Delta_A^{\bot}, \beta_1+\beta_2\in\Delta_A^{\bot}$, we then have $wt_L(c_L(v))=(q-1)q^{|B|-1}(2q^m-q^{|A|})$. On the other hand, since $\beta_1,\beta_2$ run over $\mathbb{F}_q^m$, the frequency of this Lee weight is $2(q^{m-|A|}-1)(q^m-q^{m-|A|})$.   Similarly, if $\beta_2,\beta_1+\beta_2\notin\Delta_A^{\bot}$, then $wt_L(c_L(v))=2(q-1)q^{|B|-1}(q^m-q^{|A|})$. Moreover, one has that the number of  $v$ is $(q^m-q^{m-|A|})^2$. This completes the proof.
\end{IEEEproof}

\begin{table}[!htb]
\footnotesize
\centering
\renewcommand{\arraystretch}{0.8}
\caption{Lee weight distribution in Proposition \ref{thm4}(2)}\label{Lee weightin Theorem2(2)}
\setlength{\tabcolsep}{6mm}{
\begin{tabular}{|c|c|}
\hline
 Lee weight &  Frequency\\
\hline\hline
0 & 1\\ \hline
$2(q-1)q^{|A|-1}(q^m-q^{|B|})$&$q^m-q^{m-|B|}$\\ \hline
$(q-1)q^{|A|-1}(2q^m-q^{|B|})$ &$q^{m-|A\cup B|}(q^{|A\backslash B|}-1)$\\ \hline
$2(q-1)q^{m+|A|-1}$&$q^{m-|A\cup B|}-1$\\ \hline
\end{tabular}}
\end{table}

\begin{table}[!htb]
\footnotesize
\centering
\renewcommand{\arraystretch}{0.8}
\caption{Lee weight distribution in Proposition \ref{thm4}(3)}\label{Lee weightin Theorem2(3)}
\setlength{\tabcolsep}{6mm}{
\begin{tabular}{|c|c|}
\hline
 Lee weight &  Frequency\\
\hline\hline
0 & 1\\ \hline
$(q-1)q^{m+|B|-1}$&$q^{m-|A\cup B|}-1$\\ \hline
$(q-1)q^{|B|-1}(q^m-q^{|A|})$ &$q^{m-|A\cup B|}(q^{|A\backslash B|}-1)$\\ \hline
$2(q-1)q^{|B|-1}(q^m-q^{|A|})$&$q^{m}-q^{m-|B|}$\\ \hline
\end{tabular}}
\end{table}

\begin{table}[!htb]
\footnotesize
\centering
\renewcommand{\arraystretch}{0.8}
\caption{Lee weight distribution in Proposition \ref{thm4}(4)}\label{Lee weightin Theorem2(4)}
\setlength{\tabcolsep}{6mm}{
\begin{tabular}{|c|c|}
\hline
 Lee weight &  Frequency\\
\hline\hline
0 & 1\\ \hline
$2(q-1)(q^{m-1}-q^{|A|-1})(q^m-q^{|B|})$&$q^m-q^{m-|B|}$\\ \hline
$(q-1)(q^{m-1}-q^{|A|-1})(2q^m-q^{|B|})$ &$q^{m-|A\cup B|}(q^{|A\backslash B|}-1)$\\ \hline
$(q-1)q^{m-1}(2q^{m}-2q^{|A|}-q^{|B|})$&$q^{m-|A\cup B|}-1$\\ \hline
\end{tabular}}
\end{table}
\begin{table}[!htb]
\footnotesize
\centering
\renewcommand{\arraystretch}{0.8}
\caption{Lee weight distribution in Proposition \ref{thm4}(5)}\label{Lee weightin Theorem2(5)}
\setlength{\tabcolsep}{6mm}{
\begin{tabular}{|c|c|}
\hline
 Lee weight &  Frequency\\
\hline\hline
0 & 1\\ \hline
$2(q-1)(q^{2m-1}-q^{|A|+|B|-1})$&$q^m-q^{m-|B|}$\\ \hline
$(q-1)(2q^{2m-1}-q^{|A|+|B|-1})$ &$q^{m-|A\cup B|}(q^{|A\backslash B|}-1)$\\ \hline
$2(q-1)q^{2m-1}$&$q^{m-|A\cup B|}-1$\\ \hline
\end{tabular}}
\end{table}
 The following result can be obtained from Theorem \ref{Thm2} and we omit its proof.
\begin{prop}\label{thm4}Let $s$, $m$ be  positive integers, $q=p^s$,  $\mathcal{R}=F^s$ and $A,B\subseteq[m]$ provided that $A\backslash B\ne\emptyset$. Let $\mathcal{C}_{L}$ be defined in \eqref{Ring-Trace-Code}.
\begin{enumerate}
    \item [(1)]  Let $L=a\Delta_A+c\Delta_B$. Then $\mathcal{C}_{L}$ has length
$q^{|A|+|B|}$, size $q^{ |A\cup B|}$ and
the Lee weight enumerator
 $$ 1+(q^{|A\backslash B|}-1)x^{(q-1)q^{|A|+|B|-1}}+(q^{|A\cup B|}-q^{|A\backslash B|})x^{2(q-1)q^{|A|+|B|-1}}.$$

    \item [(2)]  Let $L=a\Delta_A+c\Delta_B^{c}$ and $|B|<m$. Then  $\mathcal{C}_{L}$ has length
$q^{|A|}(q^m-q^{|B|})$, size $ q^{m}$ and
the Lee weight distribution in TABLE \ref{Lee weightin Theorem2(2)}.

    \item [(3)]  Let $L=a\Delta_A^{c}+c\Delta_B$ and $|A|<m$. Then  $\mathcal{C}_{L}$ has length
$q^{|B|}(q^m-q^{|A|})$, size $ q^{m}$ and the Lee weight distribution in TABLE \ref{Lee weightin Theorem2(3)}.

    \item [(4)]  Let $L=a\Delta_A^{c}+c\Delta_B^{c}$ and $|A|,|B|<m$. Then  $\mathcal{C}_{L}$ has length
$(q^{m}-q^{|A|})(q^m-q^{|B|})$, size $ q^{m}$ and the  Lee weight distribution in TABLE \ref{Lee weightin Theorem2(4)}.
    \item [(5)]  Let $L=(a\Delta_A+c\Delta_B)^{c}$ and $|A|+|B|<2m$. Then
$\mathcal{C}_{L}$ has length
 $q^{2m}-q^{|A|+|B|}$, size $ q^{m}$ and the  Lee weight distribution in TABLE \ref{Lee weightin Theorem2(5)}.
  \end{enumerate}
\end{prop}

\begin{remark} When $q=2$ in Propositions \ref{thm3} and \ref{thm4},  the conclusions are reduced to \cite[Theorem 10]{SS} and \cite[Theorem 14]{SS}, respectively. Therefore, the results in \cite{SS} are special cases of ours. Besides, it should be noted that when  choosing the simplicial complexes $\Delta_1$ and $\Delta_2$ with more than one element,  we can also determine the Lee weight by analyzing the relations between $(\beta_1,\beta_2)$ and $\Delta_{\cap S_1}$, $\Delta_{\cap S_2}$ case by case carefully, where $S_1\subseteq\mathcal{F}$ and $S_2\subseteq\mathcal{H}$.
\end{remark}

\section{Structures of  the Gray Image Codes and  Subfield-like Codes}
Recall that the Gray map $\phi$ in Section II is an isometry from $(\mathcal{R}^m,d_L)$ to $(\mathbb{F}_q^{2m},d_H)$.  Assume that $\mathcal{C}_L$ be given in Theorems \ref{thm1} and \ref{Thm2}. Then the parameters of the $q$-ary Gary image $\phi(\mathcal{C}_L)$ can be given as  TABLE \ref{Gray-image}. In this section, we study the  self-orthogonality, minimality and optimality of the codes $\phi(\mathcal{C}_L)$ and the subfield-like codes $\varphi(\mathcal{C}_L)$ in Theorem \ref{subfield-like}.
\begin{table}[!htb]
\footnotesize
\centering
\renewcommand{\arraystretch}{0.8}
\caption{Parameters of  $\phi(\mathcal{C}_L)$ for $\mathcal{C}_L$ in Theorems \ref{thm1} and \ref{Thm2}}\label{Gray-image}
\newcommand{\tabincell}[2]{\begin{tabular}{@{}#1@{}}#2\end{tabular}}
\begin{tabular}{|c|c|c|}
\hline
 Nos. & $\mathcal{C}_L$ as in & $[n,k,d]$  of  $\phi(\mathcal{C}_L)$ \\
\hline\hline
1&Theorem \ref{thm1}(1)&$[2 |\Delta_1| |\Delta_2|, 2|\cup_{i\in[l]}F_i|, |\Delta_2| (q-1) \mathop{\min}_{i\in [l]}q^{|F_i|-1}]$\\ \hline
2&Theorem \ref{thm1}(2)&$[2 |\Delta_1| (q^m-|\Delta_2|), 2|\cup_{i\in[l]} F_i|,(q-1)(q^m- |\Delta_2|) \mathop{\min}_{i\in [l]}q^{|F_i|-1}]$\\ \hline
3&Theorem \ref{thm1}(3)&$[2 (q^m-|\Delta_1|) |\Delta_2|, 2m,$ $ (q-1)|\Delta_2|(q^{m-1}-\sum_{i\in[l]}q^{|F_i|-1})]$\\ \hline
4&Theorem \ref{thm1}(4)&$[2(q^m- |\Delta_1|)(q^m- |\Delta_2|), 2m,$ $ (q-1)(q^m-|\Delta_2|)(q^{m-1}-\sum_{i\in[l]}q^{|F_i|-1})]$\\ \hline
5&Theorem \ref{thm1}(5)&$[ 2(q^{2m}-|\Delta_1| |\Delta_2|), 2m,$ $ (q-1)(q^{2m-1}-|\Delta_2|\sum_{i\in[l]}q^{|F_i|-1})]$\\ \hline
6&Theorem \ref{Thm2}(1)&$[2 |\Delta_1| |\Delta_2|, |(\cup_{i\in[l]}F_i)\cup(\cup_{j\in[k]}H_j)|,  (q-1) \mathop{\min}_{i\in [l],j\in[k]}\{|\Delta_2|q^{|F_i|-1},2|\Delta_1|q^{|H_j|-1}\}]$\\ \hline
7&Theorem \ref{Thm2}(2)&$[2 |\Delta_1| (q^m-|\Delta_2|), m,$ $  2(q-1)|\Delta_1|(q^{m-1}-\sum_{j\in[k]}q^{|H_j|-1})]$\\ \hline
8&Theorem \ref{Thm2}(3)&$[2(q^m- |\Delta_1|) |\Delta_2|, m,$ $\frac{q-1}{q}(q^{m}|\Delta_2|-\kappa_1)$\\ \hline
9&Theorem \ref{Thm2}(4)&$[2(q^m- |\Delta_1|)(q^m- |\Delta_2|), m, \frac{q-1}{q}(q^m(|\Delta_1^c|-|\Delta_1|+|\Delta_2^c|)-\kappa_2)]$\\ \hline
10&Theorem \ref{Thm2}(5)&$[ 2( q^{2m}-|\Delta_1| |\Delta_2|), m,$ $2(q-1)(q^{2m-1}-|\Delta_1|\sum_{j\in[k]}q^{|H_j|-1})]$\\ \hline
\end{tabular}
\makecell[t]{
 \quad \quad \quad \quad \; $\kappa_1,\kappa_2$ are defined in Theorem \ref{Thm2}(3) and Theorem \ref{Thm2}(4), respectively.}
\end{table}
\subsection{ Self-orthogonal Codes }

Let $\mathcal{C}$ be an $[n,k,d]$ linear code over $\mathbb{F}_q$. The dual of $\mathcal{C}$, which is denoted by $\mathcal{C}^\bot$, is defined as
$$\mathcal{C}^\bot=\{x\in\mathbb{F}_q^n: \langle x,c\rangle_q=0 \;\text{for all}\;c\in\mathcal{C}\}.$$
The code $\mathcal{C}$  is  self-orthogonal provided $\mathcal{C}\subseteq \mathcal{C}^\bot$. When $\mathcal{C}$ is a  quaternary code, namely, $q=4$, it is often useful to
consider another inner product called the Hermitian inner product, given by $\langle x,y\rangle_q^H=\langle x,\bar{y}\rangle_q=\sum_{i=1}^n x_i\bar{y_i}$,
where $\bar{y_i}$ is the conjugation of $y_i$. Analogous to the dual code,
we can define the Hermitian dual of the quaternary code $\mathcal{C}$ to be
$$\mathcal{C}^{\bot_H}=\{x\in\mathbb{F}_q^n: \langle x,c\rangle_q^H=0 \;\text{for all}\;c\in\mathcal{C}\}.$$
We also have  that $\mathcal{C}$  is  Hermitian self-orthogonal provided $\mathcal{C}\subseteq \mathcal{C}^{\bot_H}$.
If we have the weight distribution of $\mathcal{C}$, the following
lemma is useful to determine when  the code $\mathcal{C}$ is  (Hermitian) self-orthogonal.
\begin{lem}{\rm (\cite{HWPV})}\label{self-orth}  Let $\mathcal{C}$ be a linear code over $\mathbb{F}_q$, where $q=2,3$ or 4.

\begin{enumerate}
    \item [(1)] When $q=2$, $\mathcal{C}$ is self-orthogonal if  every codeword of $\mathcal{C}$ has weight divisible by 4.
    \item [(2)]  When $q=3$,  $\mathcal{C}$ is   self-orthogonal if and only if every codeword of $\mathcal{C}$ has weight divisible by 3.
    \item [(3)] When $q=4$, $\mathcal{C}$ is Hermitian self-orthogonal  if and only if every codeword of $\mathcal{C}$ has weight divisible by 2.
\end{enumerate}
\end{lem}

Based on the above lemma,  we have the following result.
\begin{thm}\label{structure-orthogonal} Let $m$, $l$, $k$ be positive integers,  and  $\Delta_1$, $\Delta_2$ be two simplicial complexes of $\mathbb{F}_q^m$ with $\mathcal{F}=\{F_1,\dots,F_l\}$ and $\mathcal{H}=\{H_1,\dots,H_k\}$, which are the sets of maximal elements of $\Delta_1$ and  $\Delta_2$, respectively. Let $\mathcal{C}_L$ be given in Theorems \ref{thm1} and \ref{Thm2},  $\tau_1=\min\{|\cap S_1|+|\cap S_2|-1: S_1\subseteq \mathcal{F}, S_2\subseteq \mathcal{H}\;\text{and}\;\cap S_1\ne \emptyset\}$ and  $\tau_2=\min\{|\cap S_1|+|\cap S_2|-1: S_1\subseteq \mathcal{F}, S_2\subseteq \mathcal{H}\;\text{and}\;(\cap S_1)\cup(\cap S_2)\ne\emptyset\}$. Then
\begin{enumerate}
    \item [(1)] when $q=2$, $\phi(\mathcal{C}_L)$ for $\mathcal{C}_L$ in Theorem \ref{thm1} (resp. Theorem \ref{Thm2})  is self-orthogonal if  $\tau_1\geq 2$ (resp.  $\tau_2\geq 2$);
    \item [(2)]  when $q=3$,  $\phi(\mathcal{C}_L)$ for $\mathcal{C}_L$ in Theorem \ref{thm1} (resp. Theorem \ref{Thm2}) is  self-orthogonal  if  $\tau_1\geq 1$ (resp.  $\tau_2\geq 1$);
     \item [(3)]  when $q=4$,  $\phi(\mathcal{C}_L)$ for $\mathcal{C}_L$ in Theorem \ref{thm1} (resp. Theorem \ref{Thm2}) is   Hermitian self-orthogonal  if  $\tau_1\geq 1$ (resp.  $\tau_2\geq 1$).
\end{enumerate}
\end{thm}

\begin{IEEEproof}We only give the proof for the case  $\mathcal{C}_L$ in Theorem \ref{thm1}(1) and  \ref{thm1}(3) because all the other cases can be similarly proved. Let $L=a\Delta_1+c\Delta_2$. Lemmas \ref{value-of-delta}, \ref{linear-weight} and \eqref{weight-thm1(1)} give that
\begin{eqnarray*}
wt_L(c_{L}(v))
=\sum_{\substack{\emptyset\neq S_1\subseteq\mathcal{F}\\ \emptyset\neq S_2\subseteq\mathcal{H}}}(-1)^{|S_1|+|S_2|}q^{|\cap S_1|+|\cap S_2|-1}(q-1)
\left(2-\alpha(\beta_2|\Delta_{\cap S_1})-\alpha(\beta_1+\beta_2|\Delta_{\cap S_1})\right)
\end{eqnarray*}
for any $v=a\beta_1+c\beta_2\in\mathcal{R}^m$. It is clear that if $\cap S_1=\emptyset$, which induces that $\Delta_{\cap S_1}=\{0\}$, then the term $(-1)^{|S_1|+|S_2|}q^{|\cap S_1|+|\cap S_2|-1}(q-1)$ cannot exist due to $2-\alpha(\beta_2|\Delta_{\cap S_1})-\alpha(\beta_1+\beta_2|\Delta_{\cap S_1})=0$ for any $\beta_1,\beta_2$. If $\cap S_1\ne\emptyset$, then there must be some $(\beta_1,\beta_2)\in(\mathbb{F}_q^m)^2$ such that $2-\alpha(\beta_2|\Delta_{\cap S_1})-\alpha(\beta_1+\beta_2|\Delta_{\cap S_1})\ne0$. Then the result follows from Lemma \ref{self-orth}.

 Let $L=a\Delta_1^c+c\Delta_2$ and $v=a\beta_1+c\beta_2\in\mathcal{R}^m$.  By Lemmas \ref{value-of-delta}, \ref{linear-weight} and \eqref{weight-thm1-3}, one has
\begin{eqnarray*}
wt_L(c_{L}(v))& =& (q-1)\sum_{\emptyset\neq S_2\subseteq\mathcal{H}}(-1)^{|S_2|+1}q^{m+|\cap S_2|-1}(2-\delta_{0,\beta_2}-\delta_{0,\beta_1+\beta_2})\\
&&-\sum_{\substack{\emptyset\neq S_1\subseteq\mathcal{F}\\ \emptyset\neq S_2\subseteq\mathcal{H}}}(-1)^{|S_1|+|S_2|}q^{|\cap S_1|+|\cap S_2|-1}(q-1)
\left(2-\alpha(\beta_2|\Delta_{\cap S_1})-\alpha(\beta_1+\beta_2|\Delta_{\cap S_1})\right).
\end{eqnarray*}
Since $m+|\cap S_2|-1\geq|\cap S_1|+|\cap S_2|-1$, the desired result follows by the first assertion.
\end{IEEEproof}

The following result can be obtained by an  analogous proof  to Theorem \ref{structure-orthogonal}.

\begin{thm}  Let notation be the same as Theorem \ref{structure-orthogonal} and $\varphi(\mathcal{C}_L)$  be given in Theorem \ref{subfield-like}. When $q=2$, $\varphi(\mathcal{C}_L)$  is self-orthogonal if  $\tau_1\geq 2$; when $q=3$ (resp. $q=4$), $\varphi(\mathcal{C}_L)$  is self-orthogonal (resp.  Hermitian self-orthogonal)  if $\tau_1\geq 1$.
\end{thm}

\begin{remark} In Theorem \ref{structure-orthogonal}, the self-orthogonality of the Gray image codes are characterized. Particularly, if considering $l=k=1$, that is $\mathcal{C}_L$   in Propositions \ref{thm3} and \ref{thm4}, then it is obvious $\tau_1=\tau_2=|A|+|B|-1$  due to $\mathcal{F}=\{A\}$ and $\mathcal{H}=\{B\}$. One can immediately get from Theorem \ref{structure-orthogonal} that when $q=2$, $\phi(\mathcal{C}_L)$  is self-orthogonal if  $|A|+|B|\geq 3$, which is consistent with  \cite[Proposition 12]{SS}. Besides, when $q=3$ (resp. 4),  $\phi(\mathcal{C}_L)$ is  self-orthogonal  (resp.  Hermitian self-orthogonal) if $|A|+|B|\geq 2$. Therefore, the result in \cite[Proposition 12]{SS} is a special case of Theorem  \ref{structure-orthogonal}.
\end{remark}

In what follows, we mainly focus on the optimality and minimality of the codes $\phi(\mathcal{C}_L)$ for $\mathcal{C}_L$ in Theorem \ref{Thm2} and  $\varphi(\mathcal{C}_L)$  in Theorem \ref{subfield-like}.

 \subsection{Minimal Codes}
 A  codeword $u$ in a linear code $\mathcal{C}$ over $\mathbb{F}_q$ is said to be minimal if $u$ covers only the codeword $au$ for all $a\in\mathbb{F}_q$,  but no other codewords in  $\mathcal{C}$. A linear code $\mathcal{C}$  is
said to be minimal if every  codeword in $\mathcal{C}$  is minimal. Minimal linear codes have interesting applications in secret sharing,  secure two-party computation.  Thus, searching for minimal linear codes has been an interesting research topic in coding theory. A known sufficient condition for a linear code to be minimal is due to Ashikhmin-Barg.
\begin{lem}{\rm (\cite{AB})} {\rm(\bfseries{Ashikhmin-Barg})}\label{minimality}
Let $\mathcal{C}$ be  linear  code over $\mathbb{F}_q$ with $wt_{\min}$ and   $wt_{\max}$ as the minimum and maximum Hamming weights of its non-zero codewords, respectively. If $\frac{wt_{\min}}{wt_{\max}}>\frac{q-1}{q}$, then
$\mathcal{C}$  is minimal.
\end{lem}

It should be noted that most of the linear codes over $\mathbb{F}_q$ obtained from this paper are minimal. With the help of Lemma \ref{minimality}, we characterize the minimality of  these  codes.

\begin{thm}\label{thm6} Let $\phi$ be the Gray image.  Suppose $\mathcal{C}_L$ is  as in Theorem \ref{Thm2} and  $\kappa_2$ is defined in Theorem  \ref{Thm2}.
\begin{enumerate}
    \item [(1)] If $\mathcal{C}_L$ is given in  Theorem \ref{Thm2}(2), then  $\phi(\mathcal{C}_L)$ is minimal provided $q^m>\sum_{j\in[k]}q^{|H_j|+1}$.
    \item [(2)]  If $\mathcal{C}_L$ is given in  Theorem  \ref{Thm2}(4), then $\phi(\mathcal{C}_L)$ is minimal provided  $q^m(|\Delta_1^c|+|\Delta_2^c|-|\Delta_1|)>q\kappa_2+(q-1)|\Delta_2|\sum_{i\in[l]}q^{|F_i|}$.
         \item [(3)]  If $\mathcal{C}_L$ is given in  Theorem  \ref{Thm2}(5), then $\phi(\mathcal{C}_L)$ is minimal provided  $q^{2m}>|\Delta_1|\sum_{j\in[k]}q^{|H_j|+1}$.
\end{enumerate}

\end{thm}

\begin{IEEEproof}  (1) Let $\mathcal{C}_L$ be given in  Theorem \ref{Thm2}(2). By \eqref{thm2(2)-weight}, it is clear that  $wt_{\max}\leq 2(q-1)q^{m-1}|\Delta_1|$. Since  $wt_{\min}= 2(q-1)|\Delta_1|(q^{m-1}-\sum_{j\in[k]}q^{|H_j|-1})$, we have
$$\frac{wt_{\min}}{wt_{\max}}\geq\frac{2(q-1)|\Delta_1|(q^{m-1}-\sum_{j\in[k]}q^{|H_j|-1})}{2(q-1)q^{m-1}|\Delta_1|}
=1-\frac{\sum_{j\in[k]}q^{|H_j|}}{q^m}>\frac{q-1}{q}$$
due to  $q^m>\sum_{j\in[k]}q^{|H_j|+1}$. Therefore, $\phi(\mathcal{C}_L)$ is minimal from Lemma \ref{minimality}.

(2) Let $\mathcal{C}_L$ be given in  Theorem \ref{Thm2}(4). \eqref{thm2(4)-weight} shows that for any $v=a\beta_1+c\beta_2$ with $\beta_1\ne0$,
$$
wt_L(c_L(v))=(q-1)q^{m-1}(|\Delta_1^c|+|\Delta_2^c|-|\Delta_1|)+V,
$$
where
\begin{eqnarray*}
 V&=&\Big(|\Delta_1|-|\Delta_1^c|-\frac{q}{q-1}\mathcal{A}_{\Delta_1,\beta_1}\Big)\mathcal{A}_{\Delta_2,\beta_1}
+|\Delta_2|\mathcal{A}_{\Delta_1,\beta_1} \\
&\leq&|\Delta_2|\mathcal{A}_{\Delta_1,\beta_1}
\leq|\Delta_2|(q-1)\sum\nolimits_{i\in[l]}q^{|F_i|-1}.
\end{eqnarray*}
due to $|\Delta_1|-|\Delta_1^c|-\frac{q}{q-1}\mathcal{A}_{\Delta_1,\beta_1}\leq0$ and $\mathcal{A}_{\Delta_1,\beta_1}\leq(q-1)\sum\nolimits_{i\in[l]}q^{|F_i|-1}$.
Therefore, one has $wt_L(c_L(v))\leq(q-1)q^{m-1}(|\Delta_1^c|+|\Delta_2^c|-|\Delta_1|)+|\Delta_2|(q-1)\sum\nolimits_{i\in[l]}q^{|F_i|-1}$ and the equality holds if and only if $\mathcal{A}_{\Delta_1,\beta_1}=(q-1)\sum\nolimits_{i\in[l]}q^{|F_i|-1}$ and $\mathcal{A}_{\Delta_2,\beta_1}=0$.  One can easily check that such  $\beta_1$  exists due to Lemma \ref{linear-weight}, Lemma \ref{equi-beta-set} and $F_i\backslash\big((\cup_{\ell\in[l]\backslash\{i\}}F_\ell)\cup(\cup_{t\in[k]}H_t)\big)
   \neq\emptyset$ for any $i\in[l]$. This indicates that $$wt_{\max}=(q-1)q^{m-1}(|\Delta_1^c|+|\Delta_2^c|-|\Delta_1|)+|\Delta_2|(q-1)\sum\nolimits_{i\in[l]}q^{|F_i|-1}.$$ Since
   $wt_{\min}=\frac{q-1}{q}(q^{m}(|\Delta_1^c|-|\Delta_1|+|\Delta_2^c|)-\kappa_2)$ from Theorem  \ref{Thm2}(4) and  $q^m(|\Delta_1^c|+|\Delta_2^c|-|\Delta_1|)>q\kappa_2+(q-1)|\Delta_2|\sum_{i\in[l]}q^{|F_i|}$, we have
   $$\frac{wt_{\min}}{wt_{\max}}=\frac{q^{m}(|\Delta_1^c|-|\Delta_1|+|\Delta_2^c|)-\kappa_2}
   {q^{m}(|\Delta_1^c|+|\Delta_2^c|-|\Delta_1|)+|\Delta_2|\sum\nolimits_{i\in[l]}q^{|F_i|}}=
   1-\frac{\kappa_2+|\Delta_2|\sum\nolimits_{i\in[l]}q^{|F_i|}}{q^{m}(|\Delta_1^c|+|\Delta_2^c|-|\Delta_1|)
   +|\Delta_2|\sum\nolimits_{i\in[l]}q^{|F_i|}}>1-\frac{1}{q}.$$
   Then the result follows from Lemma \ref{minimality}.
Theorem \ref{thm6}(3) can be similarly proved and thus we omit the detailed proof here. This
completes the proof.
\end{IEEEproof}

When considering the linear code $\varphi(\mathcal{C}_L)$  in Theorem \ref{subfield-like}, one can also give the sufficient condition to ensure the minimality of  $\varphi(\mathcal{C}_L)$.

\begin{thm}\label{thm7} Let $\varphi(\mathcal{C}_L)$  be given  in Theorem \ref{subfield-like}.
 The codes  $\varphi(\mathcal{C}_L)$ in Theorems \ref{subfield-like}(1) and  \ref{subfield-like}(2) are  minimal  if and only if $\Delta_1$ has exactly one maximal element. The codes  $\varphi(\mathcal{C}_L)$ in Theorems \ref{subfield-like}(3) and \ref{subfield-like}(4) are  minimal provided $q^m>\sum_{i\in[l]}q^{|F_i|+1}$. Besides, $\varphi(\mathcal{C}_L)$ in Theorem \ref{subfield-like}(5) is  minimal if  $q^{2m}>|\Delta_2|\sum_{i\in[l]}q^{|F_i|+1}$.
\end{thm}
 \subsection{Distance-Optimal Codes}
One of the important problems in coding theory is to find  linear codes over $\mathbb{F}_q$ having the largest minimum distance for  given length  and dimension. An $[n, k, d]$ code $\mathcal{C}$  is called distance optimal if no $[n, k, d+1]$ code exists, and is called almost optimal if the code $[n, k, d + 1]$ is optimal.
\begin{lem}{\rm (\cite{HWPV})} {\rm(\bfseries{Griesmer bound})}\label{Griesmer}
Let $\mathcal{C}$ be $[n,k,d]$ linear  code over $\mathbb{F}_q$, then
$$
  \sum_{i=0}^{k-1}\lceil\frac{d}{q^i}\rceil\leq n,
$$
where $\lceil\cdot\rceil$ denotes the ceiling function. $\mathcal{C}$  is called a Griesmer code if it meets the Griesmer bound with   equality.  It is well-known  that Griesmer codes are always distance-optimal, but not conversely.
\end{lem}

Let's  consider the Gray image code $ \phi(\mathcal{C}_L)$ for $\mathcal{C}_L$ in Theorem \ref{Thm2}(2) and  the subfield-like code $\varphi(\mathcal{C}_L)$  in Theorem \ref{subfield-like}(3). Numerous experiments shows that many distance-optimal codes with respect to the Griesmer bound can be derived from $ \phi(\mathcal{C}_L)$ and  $\varphi(\mathcal{C}_L)$. For example, when the   families of maximal elements of $\Delta_1$ and $\Delta_2$ are $\mathcal{F}=\{A\}$ and $\mathcal{H}=\{B\}$, respectively, by virtue of  Lemma \ref{Griesmer}, we have the following result and we omit the proof.

\begin{prop}Let $\mathcal{C}_L$ and  $\varphi(\mathcal{C}_L)$ be as in Theorem \ref{Thm2}(2) and Theorem \ref{subfield-like}(3), respectively. Let
 $\mathcal{F}=\{A\}$, $\mathcal{H}=\{B\}$ in both Theorem \ref{Thm2} and Theorem \ref{subfield-like}.
\begin{enumerate}
    \item [(1)]  Let $|A|+|B|<m$. If  $2q^{|A|}< |A|+|B|+2$ for $q=2$ or  $0<2q^{|A|}< |A|+|B|+1$ for $q>2$, then  $\phi(\mathcal{C}_L)$ is distance-optimal with respect to the Griesmer bound. If  $1\leq q^{|B|}< |A|+|B|+1$, then  $\varphi(\mathcal{C}_L)$ is distance-optimal with respect to the Griesmer bound.
      \item [(2)] Let $m\leq |A|+|B|\leq 2m-1$. If $2q^{|A|+|B|-m}(q^{m-|B|}-1)<m$ (resp.  $q^{|A|+|B|-m}(q^{m-|A|}-1)<m$),   then  $\phi(\mathcal{C}_L)$ (resp. $\varphi(\mathcal{C}_L)$) is distance-optimal with respect to the Griesmer bound.
\end{enumerate}
\end{prop}

The above conclusion is consistent with that in \cite[Theorem 13 2) and Theorem 16 3)]{SS} when $q=2$. Therefore, the result in \cite[Theorems 13 and 16]{SS} is a special case of ours. We need to mention that many other optimal linear codes can be obtained when we choose suitable simplicial complexes  $\Delta_1$ and $\Delta_2$, which may have more than  one maximal element. However, due to the uncertainty of $\Delta_1$ and $\Delta_2$, it is difficult to give the necessary and sufficient condition under which $ \phi(\mathcal{C}_L)$ and  $\varphi(\mathcal{C}_L)$ are optimal for  general  $\Delta_1$ and $\Delta_2$. When selecting $\Delta_1$ and $\Delta_2$ with certain forms, we describe the optimality of  $\varphi(\mathcal{C}_L)$ for the special case (see Proposition \ref{op1}).

\begin{prop}\label{op1}  Let $m$, $k$ be positive integers,  and  $\Delta_1$, $\Delta_2$ be two simplicial complexes of $\mathbb{F}_q^m$  which have  the sets of maximal elements $\mathcal{F}=\{F_1\}$ and $\mathcal{H}=\{H_1,\dots,H_k\}$  with $m>|F_1|$, $\min_{i\in[k]}\{|H_i|\}+|F_1|\geq   m$ and $H_i\cap H_j=\emptyset$ for any $1\leq i<j\leq k$, respectively.  Define $k-1=\sum_{j=\tau}^\ell a_jq^j$, where $a_j\in[q-1]$, $a_\tau, a_\ell>0$ and $0\leq\tau\leq\ell$.  Assume that $\varphi(\mathcal{C}_L)$  is  given in Theorem \ref{subfield-like}(3). Then   $\varphi(\mathcal{C}_L)$  is distance-optimal with respect to the Griesmer bound provided  $\tau+|F_1|>m$ and $(1-q^{|F_1|-m})|\Delta_2|<m$.
\end{prop}

\begin{IEEEproof} Note that the parameter of $\varphi(\mathcal{C}_L)$ is $[n,k,d]=[(q^m-q^{|F_1|})|\Delta_2|, m, (q-1)|\Delta_2|(q^{m-1}-q^{|F_1|-1})]$ with $|\Delta_2|=\sum_{j\in[k]}q^{|H_j|}-k+1$ due to  $H_i\cap H_j=\emptyset$ for any $1\leq i<j\leq k$. Since $\min_{i\in[k]}\{|H_i|\}+|F_1|\geq   m$ and $\tau+|F_1|>m$, we have $q^i| (k-1)q^{|F_1|-1}$ for any $i\in[m-1]$ and further,
 \begin{eqnarray*}
  \sum_{i=0}^{m-1}\Big\lceil\frac{(q-1)|\Delta_2|(q^{m-1}-q^{|F_1|-1})}{q^i}\Big\rceil &=  &\sum_{i=0}^{m-1}\frac{(q-1)|\Delta_2|(q^m-q^{|F_1|})}{q^i}\\
  &=&
  n-|\Delta_2|(1-q^{|F_1|-m})<0\\
 \end{eqnarray*}
due to $m> |F_1|$.  Besides,
 \begin{eqnarray*}
  \sum_{i=0}^{m-1}\Big\lceil\frac{(q-1)|\Delta_2|(q^{m-1}-q^{|F_1|-1})+1}{q^i}\Big\rceil =
  n-|\Delta_2|(1-q^{|F_1|-m})+m.\\
 \end{eqnarray*} The result follows from  Lemma \ref{Griesmer} immediately.
This completes the proof.
\end{IEEEproof}

\begin{remark}We claim that there always exist $\mathcal{F}$ and $\mathcal{H}$ meeting the assumption in proposition \ref{op1} for any $q$. In particular, letting $|F_1|=m-1$, $k=q^2+1$ and  $H_i\cap H_j\ne\emptyset$ for any $1\leq i<j\leq k$. It is clear that $\tau+|F_1|=m+1>m$  and $(1-q^{|F_1|-m})|\Delta_2|=(q-1)(\sum_{i\in[ k]}q^{|H_j|-1}-q)$. When $m$ is large enough, one can obtain that $(1-q^{|F_1|-m})|\Delta_2|<m$, which indicates that  the optimal codes  in proposition \ref{op1} exist  for any prime power $q$.
\end{remark}

\begin{remark} This section investigates the structures of the Gray image codes and the subfield-like codes. The minimality, self-orthogonality and optimality of these codes are characterized. Eventually, infinite families of minimal codes and optimal codes can be obtained. On the other hand, one can find in Section VI that new codes can be derived in our construction. Therefore,  these results  give an affirmative answer to  the second question in Introduction B.
\end{remark}

\begin{problem} Let  the Gray image code be $ \phi(\mathcal{C}_L)$ for $\mathcal{C}_L$ in Theorem \ref{Thm2}(2) and  the subfield-like code be  $\varphi(\mathcal{C}_L)$  in Theorem \ref{subfield-like}(3). Determining the necessary and sufficient conditions  on $\Delta_1$ and $\Delta_2$ under which $\phi(\mathcal{C}_L)$ or $\varphi(\mathcal{C}_L)$ is optimal.
\end{problem}
\section{Comparisons  and Conclusion Remarks}
Let $\mathcal{R}$ be any extension of the non-unital non-commutative rings $E$ and $F$. In this paper, we constructed several infinite families codes over $\mathcal{R}$ using  simplicial complexes which have arbitrary number of maximal elements.  The principal parameters of these codes were given,  and the Lee weight distributions  were characterized explicitly when   the applied simplicial complexes generated by one maximal element.   Besides, we studied the corresponding subfield-like linear codes and the Gray image linear codes.  Consequently,  infinite families of minimal codes, distance-optimal codes and self-orthogonal codes were produced by analyzing the structures of these linear codes. To show significant advantages of our codes, in this section, we present two tables. In Table \ref{Known codes}, we list some known  codes over rings constructed from simplicial complexes for the convenience of the reader. Compared with known results, the codes obtained in this work have flexible and new parameters. Table \ref{Distance-optimal  codes} presents some examples of distance-optimal  linear codes  from the subfield-like linear codes and the Gray image linear codes in small dimensions. To the best of our knowledge, this is the first paper to obtain codes over the extension rings of  non-unital,  non-commutative rings  by using general simplicial complexes of $\mathbb{F}_q^m$ for any prime power $q$. It is worth noting that the technique in this paper can be employed to
 extend the previous results of \cite{LShi}, \cite{SS2}, \cite{SL2}, \cite{WZY},
where the  utilized simplicial complexes of $\mathbb{F}_2^m$
have one maximal element.

\begin{table}[!htb]
\footnotesize
\centering
\renewcommand{\arraystretch}{0.6}
\caption{Some known codes over ring $a\mathbb{F}_q+c\mathbb{F}_q$ using simplicial complexes}\label{Known codes}
\newcommand{\tabincell}[2]{\begin{tabular}{@{}#1@{}}#2\end{tabular}}
\begin{tabular}{|c|c|c|c|c|}
\hline
$q$ & Values of $a,c $ &Defining sets & Parameters ($n,|\mathcal{C}_L|,\min d_L$)  &  References \\
\hline\hline
 \multirow{2}{*}{$q=2$}   &$a=1,c^2=0$  &     $\mathbb{F}_2^m+c\Delta_A$ &  $(2^{m+|A|},2^{2m},2^{m+|A|-1})$   &    \multirow{2}{*}{\cite{LShi}} \\  \cline{3-4}
                                                                                           &  &   $\mathbb{F}_2^m+c\Delta_A^c$ & $(2^{m}(2^m-2^{|A|}),2^{2m},2^{m}(2^m-2^{|A|-1}))$   &  \\ \hline

   \multirow{2}{*}{$q=2$}   & \multirow{2}{*}{$a=1,c^2=0$}   &     $\Delta_A+c\Delta_B^c$ &  $(2^{|A|}(2^m-2^{|B|}),2^{m+|A|},2^{|A|}(2^m-2^{|B|}))$  &    \multirow{2}{*}{\cite{WZY}} \\  \cline{3-4}
                                                                                              &  &    $\Delta_A^c+c\Delta_B^c$ & $((2^m-2^{|A|})(2^{m}-2^{|B|}),2^{2m},(2^m-2^{|A|})(2^{m}-2^{|B|})-2^{|A|+|B|})$   &  \\ \hline

      \multirow{6}{*}{$q=2$}   & \multirow{6}{*}{$a^2=c,ac=0$}   &     $a\Delta_A+c\Delta_B$ &  $(2^{|A|+|B|},2^{|A|},2^{|A|+|B|})$   &    \multirow{6}{*}{\cite{SS2}} \\ \cline{3-4}
                                                                                               &  &    $a\Delta_A^c+c\Delta_B$& $((2^m-2^{|A|})2^{|B|},2^{m},(2^m-2^{|A|})2^{|B|})$  &  \\  \cline{3-4}
                                                                                                &  &    $a\Delta_A+c\Delta_B^c$& $((2^m-2^{|B|})2^{|A|},2^{|A|},(2^m-2^{|B|})2^{|A|})$  &  \\  \cline{3-4}
                                                                                                &  &    $a\Delta_A^c+c\Delta_B^c$ & $((2^m-2^{|A|})(2^m-2^{|B|}),2^{m},(2^m-2^{|A|})(2^m-2^{|B|}))$  &  \\ \cline{3-4}
                                                                                             &  &    $(a\Delta_A+c\Delta_B)^c$ & $(2^{2m}-2^{|A|+|B|},2^{m},2^{2m}-2^{|A|+|B|})$   &  \\ \hline

       \multirow{13}{*}{$q=2$}   & \multirow{6}{*}{$a^2=a,ac=0$}   &     $a\Delta_A+c\Delta_B$ &  $(2^{|A|+|B|},2^{2|A|},2^{|A|+|B|-1})$   &    \multirow{13}{*}{\cite{SS}} \\ \cline{3-4}
                                                                                               &  &    $a\Delta_A^c+c\Delta_B$ & $((2^m-2^{|A|})2^{|B|},2^{2m},2^{m+|B|-1}-2^{|A|+|B|-1})$  &  \\ \cline{3-4}
                                                                                                &  &    $a\Delta_A+c\Delta_B^c$ & $((2^m-2^{|B|})2^{|A|},2^{2|A|},2^{|A|-1}(2^m-2^{|B|}))$  &  \\  \cline{3-4}
                                                                                                &  &    $a\Delta_A^c+c\Delta_B^c$ & $((2^m-2^{|A|})(2^m-2^{|B|}),2^{2m},(2^{m-1}-2^{|A|-1})(2^m-2^{|B|}))$  &  \\ \cline{3-4}
                                                                                             &  &    $(a\Delta_A+c\Delta_B)^c$ & $(2^{2m}-2^{|A|+|B|},2^{2m},2^{2m-1}-2^{|A|+|B|-1})$   &  \\   \cline{2-4}
       & \multirow{6}{*}{$a^2=a,ac=c$}   &     $a\Delta_A+c\Delta_B$ &  $(2^{|A|+|B|},2^{|A|\cup |B|},2^{|A|+|B|-1})$,   &   \\  \cline{3-4}
                                                                                                 &  &    $a\Delta_A^c+c\Delta_B$ & $((2^m-2^{|A|})2^{|B|},2^{m},2^{|B|-1}(2^{m}-2^{|A|}))$   &  \\ \cline{3-4}
                                                                                                &  &    $a\Delta_A+c\Delta_B^c$ & $((2^m-2^{|B|})2^{|A|},2^{m},2^{|A|}(2^m-2^{|B|}))$  &  \\ \cline{3-4}
                                                                                                &  &    $a\Delta_A^c+c\Delta_B^c$ & $((2^m-2^{|A|})(2^m-2^{|B|}),2^{m},(2^{m}-2^{|A|})(2^m-2^{|B|}))$   &  \\ \cline{3-4}
                                                                                             &  &    $(a\Delta_A+c\Delta_B)^c$ & $(2^{2m}-2^{|A|+|B|},2^{m},2^{2m}-2^{|A|+|B|})$   &  \\  \hline

              \multirow{18}{*}{$q$ any}   & \multirow{8}{*}{$a^2=a,ac=0$}   &     $a\Delta_1+c\Delta_2$ &  $( |\Delta_1||\Delta_2|, q^{2|\cup_{i\in[l]}F_i|},|\Delta_2| (q-1) \mathop{\min}_{i\in [l]}q^{|F_i|-1} )$   &    \multirow{18}{*}{This paper} \\ \cline{3-4}
                                                                                               &  &    $a\Delta_1^c+c\Delta_2$ & $( (q^m- |\Delta_1|) |\Delta_2|,q^{2m},(q-1)|\Delta_2|(q^{m-1}-\sum_{i\in[l]}q^{|F_i|-1}))$  &  \\ \cline{3-4}
                                                                                                &  &    $a\Delta_1+c\Delta_2^c$ & $( |\Delta_1| (q^m-|\Delta_2|), q^{2|\cup_{i\in[l]}F_i|},(q^m-|\Delta_2|) (q-1) \mathop{\min}_{i\in [l]}q^{|F_i|-1})$  &  \\  \cline{3-4}
                                                                                                &  &    $a\Delta_1^c+c\Delta_2^c$ & \tabincell{c}{ $((q^m- |\Delta_1|)(q^m- |\Delta_2|),q^{2m},$\\
                                                                                                $ (q-1)(q^m- |\Delta_2|)(q^{m-1}-\sum_{i\in[l]}q^{|F_i|-1}))$ } &  \\ \cline{3-4}
                                                                                             &  &    $(a\Delta_1+c\Delta_2)^c$ & $( q^{2m}-|\Delta_1| |\Delta_2|,q^{2m},(q-1)(q^{2m-1}-|\Delta_2|\sum_{i\in[l]}q^{|F_i|-1}))$   &  \\   \cline{2-4}
       & \multirow{8}{*}{$a^2=a,ac=c$}   &     $a\Delta_1+c\Delta_2$ & \tabincell{c}{ $(|\Delta_1| |\Delta_2|,q^{ |(\cup_{i\in[l]}F_i)\cup(\cup_{j\in[k]}H_j)|},$\\
       $(q-1) \mathop{\min}_{i\in [l],j\in[k]}\{|\Delta_2|q^{|F_i|-1},2|\Delta_1|q^{|H_j|-1}\})$ } &   \\  \cline{3-4}
                                                                                                 &  &    $a\Delta_1^c+c\Delta_2$ & $((q^m- |\Delta_1|) |\Delta_2|,q^m,\frac{q-1}{q}(q^m|\Delta_2|-\kappa_1))$   &  \\ \cline{3-4}
                                                                                                &  &    $a\Delta_1+c\Delta_2^c$ & $( |\Delta_1|(q^m- |\Delta_2|),q^m,2(q-1)|\Delta_1|(q^{m-1}-\sum_{j\in[k]}q^{|H_j|-1}))$  &  \\ \cline{3-4}
                                                                                                &  &    $a\Delta_1^c+c\Delta_2^c$ & $((q^m- |\Delta_1|)(q^m- |\Delta_2|),q^m,\frac{q-1}{q}(q^{m}(|\Delta_1^c|-|\Delta_1|+|\Delta_2^c|)-\kappa_2))$   &  \\ \cline{3-4}
                                                                                             &  &    $(a\Delta_1+c\Delta_2)^c$ & $2(q-1)(q^{2m-1}-|\Delta_1|\sum_{j\in[k]}q^{|H_j|-1})$   &  \\  \hline

\end{tabular}
\makecell[l]{
\quad 1. $A,B\subseteq[m]$, $\Delta_1,\Delta_2$ are generated by the sets of  maximal elements $\mathcal{F}=\{F_1,\dots,F_l\}$ and $\mathcal{H}=\{H_1,\dots,H_k\}$, respectively.\\
\quad 2. $|\Delta_1|=\sum_{\emptyset\neq S\subseteq\mathcal{F}}(-1)^{|S|+1}q^{|\cap S|}$, $|\Delta_2|=\sum_{\emptyset\neq S\subseteq\mathcal{H}}(-1)^{|S|+1}q^{|\cap S|}$.}
\end{table}

\begin{table}[!htb]
\footnotesize
\centering
\renewcommand{\arraystretch}{0.8}
\caption{Distance-optimal linear codes over $\mathbb{F}_q$ in small dimensions constructed in this paper}\label{Distance-optimal  codes}
\begin{tabular}{|c|c|c|c|c|c|}
\hline
$q$ & $m$  & $\mathcal{F}$ & $\mathcal{H}$ &$[n,k,d]$  &  Results \\
\hline\hline
\multirow{10}{*}{2} &3 &$\{\{2\}\}$  &$\{\{1,2\}\}$  &$[16,3,8]$  & \multirow{10}{*}{Theorem \ref{Thm2}}\\ \cline{2-5}
 &\multirow{2}{*}{4} &$\{\{1\}\}$  &$\{\{1,3,4\}\}$  &$[32,4,16]$  & \\ \cline{3-5}
 & &$\{\{1\}\}$  &$\{\{1,2\}\}$  &$[48,4,24]$  & \\  \cline{2-5}

 &\multirow{2}{*}{5} &$\{\{2\}\}$  &$\{\{1,2,3,4\}\}$  &$[64,5,32]$  & \\  \cline{3-5}

 & &$\{\{1\}\}$  &$\{\{2,4\}\}$  &$[112,5,56]$  & \\  \cline{2-5}

  &\multirow{5}{*}{6} &$\{\{1\}\}$  &$\{\{1,2,3,4,5\}\}$  &$[128,6,64]$  & \\  \cline{3-5}
 &  &$\{\{4\}\}$  &$\{\{1,3,4,5\}\}$  &$[192,6,96]$  & \\  \cline{3-5}
 & &$\{\{4\}\}$  &$\{\{2,4,5\}\}$  &$[224,6,112]$  & \\  \cline{3-5}
 & &$\{\{1\}\}$  &$\{\{3,4\}\}$  &$[240,6,120]$  & \\  \cline{3-5}
 & &$\{\{2,4\}\}$  &$\{\{1,2,4,5,6\}\}$  &$[256,6,128]$  & \\ \cline{1-5}
 $3$& 5& $\{\{3\}\}$  &$\{\{1,2,3,4\}\}$  &$[972,5,648]$  &\\ \hline

 \multirow{6}{*}{2} &3 &$\{\{1,2\}\}$  &$\{\{1\},\{3\}\}$  &$[12,3,6]$  & \multirow{10}{*}{Theorem \ref{subfield-like}}\\ \cline{2-5}
 &4 &$\{\{2,3,4\}\}$  &$\{\{3,4\},\{2\}\}$  &$[40,4,20]$  &\\ \cline{2-5}
 &\multirow{2}{*}{5}&$\{\{1,2,4,5\}\}$  &$\{\{1,4\},\{3\},\{5\}\}$  &$[96,5,48]$  &\\ \cline{3-5}
 &&$\{\{1,3,4,5\}\}$  &$\{\{1,3\},\{1,2\},\{2,5\}\}$  &$[128,5,64]$  &\\ \cline{2-5}
 &\multirow{2}{*}{6}&$\{\{1,2,3,5,6\}\}$  &$\{\{3,4\},\{2\},\{3,5\},\{4,6\}\}$  &$[288,6,144]$  &\\ \cline{3-5}
 &&$\{\{2,3,4,5,6\}\}$  &$\{\{1,6\},\{1,2\},\{1,5\},\{3,5\}\}$  &$[320,6,160]$  &\\ \cline{1-5}

  \multirow{4}{*}{3} &3  &$\{\{1,3\}\}$  &$\{\{1\}\}$  &$[54,3,36]$  &\\ \cline{2-5}
  &\multirow{2}{*}{4}&$\{\{1,2,3\}\}$  &$\{\{4\}\}$  &$[162,4,108]$  &\\ \cline{3-5}
  & &$\{\{1,4\}\}$  &$\{\{2\}\}$  &$[216,4,144]$  &\\ \cline{2-5}
  &5&$\{\{1,2,3,5\}\}$  &$\{\{2\}\}$  &$[486,5,324]$  &\\\hline
\end{tabular}
\end{table}

\section*{Acknowledgment}
This work was supported by the  Postdoctoral Fellowship Program of CPSF (No. GZB20230815), the National Natural Science Foundation of China (Nos. 12201356,  62072162, 62372445, 62032009), the Natural Science Foundation of Shandong Province of China (No. ZR2022QA048), the Natural Science Foundation of Hubei Province of China (No. 2021CFA079), the Knowledge Innovation Program of Wuhan-Basic Research (No. 2022010801010319),  the National Key Research and Development Program of China (Nos. 2021YFA1000600, 2020YFA0712300) and the Innovation Group Project of the Natural Science Foundation of Hubei Province of China (No. 2023AFA021).

\end{document}